\documentclass[aps,prx,amsmath,amssymb,amsfonts,nofootinbib,superscriptaddress,notitlepage,reprint,longbibliography,floatfix
]{revtex4-2}

\usepackage[english]{babel}
\usepackage[T1]{fontenc}
\usepackage[utf8]{inputenc} 
\usepackage[version=4]{mhchem}
\usepackage{geometry, units, textcomp, amsmath, graphicx, xcolor, lipsum, xfrac, comment, upgreek, appendix, physics}

\newcommand{\Er}{Er$^{3+}$\,}
\newcommand{\Nb}{$^{93}$Nb\,}
\newcommand{\W}{$^{183}$W\,}
\newcommand{\Ca}{$\mathrm{CaWO}_4$\,}
\renewcommand{\thefootnote}{\fnsymbol{footnote}}

\begin{document}

\author{J. Travesedo$^{1,\dagger}$, Z. W. Huang$^{1,\dagger}$, L. Mykolyshyn$^1$, N. Thill$^1$, L. Pallegoix$^1$, P. Goldner$^2$, T. Chaneliere$^3$, S. Bertaina$^4$, T. Charpentier$^5$, D. Est\`eve$^1$, P. Abgrall$^1$, D. Vion$^1$, J. O'Sullivan$^{1}$, E. Flurin$^1$, P. Bertet}

\email{emmanuel.flurin@cea.fr}
\email{patrice.bertet@cea.fr}

\affiliation{Quantronics Group, Universit\'e Paris-Saclay, CEA, CNRS, SPEC, 91191 Gif-sur-Yvette Cedex, France\\
$^2$Chimie ParisTech, PSL University, CNRS, Institut de Recherche de Chimie Paris, 75005 Paris, France\\$^3$Universit\'e Grenoble Alpes, CNRS, Grenoble, France \\
$^4$Aix-Marseille Univ. University of Toulon, IM2NP, 13013, Marseille, France \\
$^5$NIMBE, CEA, 91191 Gif-sur-Yvette Cedex, France
}




\title{Precision hyperfine spectroscopy of an individual nuclear-spin-9/2}



\begin{abstract}

\textbf{
Single-spin magnetic resonance spectroscopy promises to yield structural and chemical information at the level of individual atoms or molecules, in a non-invasive way~\cite{budakian_roadmap_2024}. Here, we use an \Er paramagnetic center in a \Ca crystal, detected by microwave photon counting at $10$\,mK~\cite{albertinale_detecting_2021,wang_single-electron_2023}, as a nanoscale magnetic sensor to measure the NMR spectrum of a proximal individual nuclear-spin-9/2 \Nb impurity with Hertz spectral resolution. From these measurements, we determine the \Nb insertion site, its position relative to the \Er, and its complete quadrupolar tensor. We moreover harness the high spectral resolution of our measurements to establish the presence of two previously unobserved terms in the spin Hamiltonian. The first describes a coupling between the \Er spin and the \Nb nuclear quadrupole; it possibly originates from a spin-dependent electrostatic interaction between the two systems. The second is a nuclear hexadecapolar term~\cite{doering_search_1986}, and may be caused by the coupling of the electric field third derivative to the \Nb nuclear hexadecapolar moment. 
}

\end{abstract}

\maketitle
\def\thefootnote{$\dagger$}\footnotetext{Both authors contributed equally to this work}\def\thefootnote{\arabic{footnote}}

The interaction between an electron and a nuclear spin in matter is at heart of magnetic resonance spectroscopy~\cite{schweiger_principles_2001} and spin-based quantum computing. This simple system is generally well described by a spin Hamiltonian that is the sum of the electron and nuclear Zeeman energies, of the magnetic hyperfine interaction, and of the nuclear electric quadrupole, involving products of electron and nuclear spin operators with power no greater than $2$. Higher-order terms are however theoretically possible~\cite{schwartz_theory_1955}; their study is of fundamental interest as it may bring new information about the nuclear structure. These terms are expected to be at least $\sim 10^5$ times weaker than the leading-order terms; hence, high spectral resolution ($\sim $Hz) is required to resolve their contribution. 

Such high spectral resolution is reached in atomic physics experiments using atomic or molecular beams, or single trapped ions, probed by combined microwave and laser excitations. This allowed measuring the third-order nuclear magnetic octupole interaction for several nuclei~\cite{jaccarino_PhysRev.94.1798,gerginov_observation_2003,lewty_spectroscopy_2012}, whereas the fourth-order nuclear electric hexadecapole interaction has not been conclusively measured so far for any nucleus. In the solid state, on the other hand, measurements of coupled electron-nuclear spin ensembles by magnetic resonance methods generally offer a lower spectral resolution (kHz at best~\cite{sielaff_pulsed_2025}), particularly when high-spin nuclei are involved. Indeed, the lines are broadened due to a combination of spin-spin interactions and of inhomogeneous distributions of the electron gyromagnetic tensor and/or of the nuclear quadrupole couplings. This precluded so far the observation of terms beyond the leading-order magnetic dipole and electric quadrupole by magnetic resonance, despite active research~\cite{segel_nuclear_1978,doering_search_1986,liao_nuclear_1994}.

Major progress in spectral resolution was brought by recent experiments carried on individual electron spins in ultra-pure crystals at low temperature, opening the way to precision hyperfine spectroscopy in the solid-sate. The hyperfine and quadrupolar interactions of $^{13}\mathrm{C}$ and $^{14}\mathrm{N}$ nuclear spins in diamond were measured with hertz resolution by a Nitrogen-Vacancy (NV) center spin, using optical detection of the NV at 4K~\cite{abobeih_atomic-scale_2019, degen_creation_2021, van_de_stolpe_mapping_2024,breitweiser_quadrupolar_2024,bartling_universal_2025}. Hertz resolution was also reported for the spin Hamiltonian parameters of $^{31}\mathrm{P}$ and $^{123}\mathrm{Sb}$ nuclear spin ($I=7/2$) of an electron spin donor in $^{28}\mathrm{Si}$-enriched silicon, using spin-to-charge conversion of the donor at 10mK for detection~\cite{fernandez_de_fuentes_navigating_2024,vaartjes2025maximizingnondemolitionnaturequantum}. 

\begin{figure*}[t]
    \includegraphics[width=\textwidth]{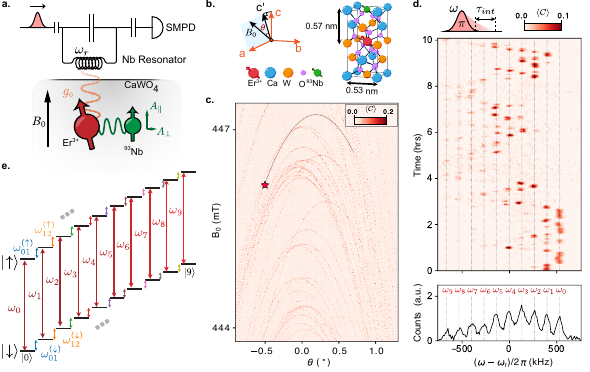}
    \caption{\label{fig1} \textbf{Schematics of the experimental setup and high-resolution electron spin spectroscopy.}   
    \textbf{a.} Sample Schematics. A superconducting resonator, fabricated on the surface of a CaWO$_4$ substrate, is magnetically coupled to a single \Er spin (red) with strength $g_0$, in turn coupled via hyperfine interaction ($A_\parallel$ and $A_\perp$ in dark green) to a \Nb nuclear spin (green). The \Er spin is excited with a microwave pulse (red) and its microwave spontaneous emission is routed towards the input of a Single Microwave Photon Detector (SMPD). A static magnetic field $B_0$ is used to tune the transition frequency of the \Er spin to the frequency of the resonator $\omega_\mathrm{r}$.
    \textbf{b.} \Ca crystal unit cell with the \Er and \Nb impurities in their assigned sites (see main text). The magnetic field $B_0$ is applied at an angle $\theta$ with respect to c', the projection of the c-axis onto the resonator plane (blue rectangle).
    \textbf{c.} Single-\Er-resolved spectroscopy. A microwave pulse of duration 15~\textmu s is applied at each $B_0$ and $\theta$, and the subsequent counts detected by the SMPD are summed during 1ms.
    The \Er spin studied is marked by a black dashed line. Most data reported in this work are obtained at $\theta = -0.5^\circ$ (red star).
    \textbf{d.} Spectroscopy of the \Er-\Nb coupled system as a function of time. Each sequence consists of applying a Gaussian pulse of duration 40 \textmu s and frequency $\omega$, and summing the SMPD counts (after a waiting time $\tau_{\text{w}}=120$~\textmu s) for $\tau_{int}=2$~ms yielding $C$. The frequency $\omega$ is swept over 1.5 MHz around $\omega_r$, with the SMPD frequency tuned to $\omega$ (see App.\ref{app:sample_details}). The sweep is repeated over $10$ hours. The top panel shows the number of counts averaged over groups of $1000$ consecutive sweeps as a function of $\omega$ and time. The bottom panel shows the number of counts averaged over the complete sweep, as a function of $\omega$. Gray dashed lines have been plotted as a guide for the eye for each EPR transition with frequencies $\omega_{n}$. \textbf{e.} Energy diagram of the \Er - \Nb coupled system. The nuclear spin states are labeled as $\ket{n}$, with $n$ varying from $0$ to $9$. The $10$ EPR-allowed transitions at $\omega_n$ are labeled and marked with red arrows. NMR transitions when the \Er is in its ground (resp. excited) state at $\omega_{n,n+1}^{(\downarrow)}$ (resp. $\omega_{n,n+1}^{(\uparrow)}$) are labeled and marked with multicolored arrows.
    }
\end{figure*}

Here, we measure the transition frequencies between the energy levels of an individual $I=9/2$ nuclear spin impurity coupled to an individual \Er electron spin in a \Ca crystal using microwave photon counting at $10$\,mK (see Fig.\ref{fig1}a). We identify the nuclear spin as a \Nb atom, determine its position relative to the \Er spin, and measure its complete quadrupolar tensor. Owing to the high spectral resolution of our measurements ($30$\,Hz when the \Er is in its excited state, $1$\,Hz when it is in its ground state, and $10$\,mHz in differential frequency measurements), we establish the existence of two previously unobserved high-order terms in the spin Hamiltonian. The first term is proportional to $S_z I_z^2$ with $S_z$ (resp. $I_z$) being the \Er (resp. \Nb) spin operator. This spin-dependent quadrupole is attributed in part to a spin-dependent electrostatic interaction between the \Er spin and the \Nb nucleus. The second term is proportional to $I_z^4$, and therefore describes a nuclear hexadecapolar interaction. Further work is needed to establish whether it is due to the coupling of the electric field third derivative to the \Nb nuclear hexadecapolar moment, which would then be probed for the first time.

The coupled \Nb -- \Er spin system was discovered fortuitously in a \Ca sample that was used in previous experiments to demonstrate polarization, spectroscopy, and read-out of individual nuclear spins of $^{183}\mathrm{W}$ \cite{travesedo_all-microwave_2025}. CaWO$_4$ is a crystal with a low nuclear magnetic moment density, since the most abundant isotope with a nuclear spin is $^{183}$W, with 14.4\% abundance, a spin 1/2, and a low magnetic moment of $0.11778 \, \mu_\text{N}$ ($\mu_N$ being the nuclear magneton), resulting in particularly long electron~\cite{le_dantec_electron_2022} and nuclear~\cite{osullivan_individual_2024} spin coherence times. 

\Er ions enter in \Ca by substitution of Ca$^{2+}$ as shown in Fig.\ref{fig1}b. The 16-fold degenerate $J=15/2$ ground state splits into eight Kramers' doublets due to the crystal field (see App.B for more details). At cryogenic temperatures, only the lowest doublet is occupied; it thus behaves as an effective $S=1/2$ spin, with an anisotropic gyromagnetic tensor $\gamma_{\mathrm{Er}}$ axially symmetric around the $c$ axis of the crystal. On top of the crystal, a superconducting thin-film niobium resonator is fabricated with frequency $\omega_\mathrm{r}/2\pi = 7.7492$ GHz and linewidth $\kappa/2\pi$ = 740 kHz. The resonator has a 300~nm-wide nanowire constriction approximately parallel to the crystalline $c$-axis (further details can be found in App.\ref{app:sample_details}). The resonator sustains an oscillating magnetic field $B_1$, whose quantum fluctuations couple to \Er ions with a strength $g_0$~\cite{bienfait_controlling_2016,albertinale_detecting_2021,wang_single-electron_2023} (see Fig.~\ref{fig1}a). A magnetic field $\mathbf{B_0}$, of magnitude $B_0$, is applied in the resonator plane at an angle $\theta$ from the $c$-axis projection on this plane (see Fig.~\ref{fig1}b). The field tunes the \Er spin frequency $\omega_S = ||\mathbf{\gamma_{\text{Er}}} \cdot \mathbf{B_0} ||$ by the Zeeman effect. When resonant with an \Er spin, the resonator performs two functions. It allows one to drive the \Er spin using microwave pulses \cite{wang_single-electron_2023} and it enhances its radiative decay rate $\Gamma_\mathrm{R} = 4 g_0^2 / \kappa$ via the Purcell effect \cite{bienfait_controlling_2016}. The emitted microwave photons are directed towards the input of a transmon-based Single Microwave Photon Detector (SMPD) \cite{lescanne_irreversible_2020, pallegoix_enhancing_2025}. 

The sample hosts a large number of addressable single \Er spins, as can be seen from the rotation pattern shown in Fig.\ref{fig1}c where each spin appears as a narrow peak in the number of detected counts $\langle C \rangle$ following an excitation pulse at $\omega_r$~\cite{wang_single-electron_2023}. To study the nuclear spin environment of one of these \Er spins, we set $B_0$ and $\theta$ on the corresponding peak, and we repeatedly measure the \Er spectrum. Strongly coupled nuclear spins are revealed by sudden jumps of the \Er resonance, due to the surrounding nuclear spin state change induced by nuclear-spin-flipping relaxation events of the \Er spin, which are weakly allowed by the transverse term in the hyperfine interaction~\cite{travesedo_all-microwave_2025}. In most studied ions, the time traces show two \Er resonances, corresponding to the coupling to one  spin-1/2 nucleus, which can be identified as a $^{183}$W by spectroscopy~\cite{travesedo_all-microwave_2025}. Here, we concentrate on one \Er spin marked by a black dashed line, and we measure it for $\theta = -0.5 ^\circ$ (see red star in Fig.~\ref{fig1}c) for most of the remaining of this work. Its time trace shows ten evenly-spaced resonances with vastly different cross-relaxation lifetimes (see Fig.\ref{fig1}d), indicating the coupling to an unknown spin-9/2 nuclear spin impurity.

To leading-order, the Hamiltonian of this coupled spin system is

\begin{equation} \label{eq:SpinHam}
    \mathcal{H} = \mathcal{H}_{eZ} + \mathcal{H}_{nZ} + \mathcal{H}_{hf} + \mathcal{H}_Q,
\end{equation}

\noindent where $\mathcal{H}_{eZ} = \omega_S S_{z}$ is the \Er spin Zeeman energy, and $\mathcal{H}_{nZ} = \omega_I I_z$ is the Zeeman energy of the nuclear spin of Larmor frequency $\omega_I = - \gamma_I B_0$, $\gamma_I$ being its gyromagnetic ratio. $\mathcal{H}_{hf} = S_z (A_\parallel I_z + A_\perp I_x)$ is the hyperfine interaction in the secular approximation. Finally, $\mathcal{H}_Q = I \cdot Q \cdot I$ is the quadrupolar term describing the nucleus electrostatic energy in the electric field gradient caused by the surrounding charges. In its principal axis basis $X,Y,Z$, it can be written as $\mathcal{H}_Q=Q_X I_X^2 + Q_Y I_X^2 + Q_Z I_X^2$ (with $Q_X + Q_Y + Q_Z = 0$), or as $\mathcal{H}_Q = \frac{C_q}{4 I (2I-1)}\cdot [3I_X^2 - I(I+1) + \eta (I_Y^2 - I_Z^2)]$, which defines the quadrupolar interaction strength $C_q$ and the biaxiality parameter $\eta$. Hamiltonian Eq.\ref{eq:SpinHam} can be re-written as

\begin{align}
\begin{split}
    \mathcal{H} = &\ket{\downarrow}\bra{\downarrow}\otimes \mathcal{H}^{(\downarrow)} + \ket{\uparrow}\bra{\uparrow}\otimes \mathcal{H}^{(\uparrow)} \\
    \mathcal{H}^{(\downarrow)}/\hbar &= -\omega_S / 2 + \omega_I^{(\downarrow)} I_z^{(\downarrow)} + \mathbf{I}\cdot \bar{\bar{Q}}^{(\downarrow)}\cdot \mathbf{I} \\ 
    \mathcal{H}^{(\uparrow)}/\hbar &= +\omega_S / 2 + \omega_I^{(\uparrow)}I_z^{(\uparrow)} + \mathbf{I}\cdot \bar{\bar{Q}}^{(\uparrow)} \cdot\mathbf{I}.
\end{split}
\label{eq2}
\end{align}

\noindent  $\mathcal{H}^{(m_S)}$ is the nuclear spin Hamiltonian when the \Er spin is in $m_S = \downarrow$ or $m_S = \uparrow$. As a result of the hyperfine interaction, the nuclear spin quantization axis $z^{(m_S)}$ and Larmor frequency $\omega_I^{(m_S)} \simeq \omega_I + m_S A_\parallel/2 $ depend on the \Er spin state (see App.\ref{app:spin_hamiltonian}). According to Hamiltonians Eq.~\ref{eq:SpinHam} and Eq.~\ref{eq2}, the quadrupolar tensor should not depend on $m_S$; here, we test this property by separately measuring $\bar{\bar{Q}}^{(\downarrow)}$ and $\bar{\bar{Q}}^{(\uparrow)}$ for each \Er spin orientation.

Fig.\ref{fig1}e shows the $20$ energy levels of the coupled spins, grouped in the two $m_S = \downarrow$ or $m_S = \uparrow$ manifolds. The nuclear spin eigenstates are mixed by the quadrupolar interaction and are therefore labeled as $\ket{n}$ in increasing order of energy, with $n$ varying between $0$ and $9$. Ten transitions between $\ket{\downarrow, n}$ and $\ket{\uparrow, n}$ at $\omega_n$ are EPR-allowed. Transitions between $\ket{m_S, n}$ and $\ket{m_S, n+1}$ at frequencies $\omega_{n,n+1}^{(m_S)} $ are NMR-allowed. Double- or zero-quantum transitions between $\ket{\downarrow, n}$ and $\ket{\uparrow, n \pm 1}$ are weakly authorized because of the slight difference of nuclear spin orientation when the \Er spin is $\ket{\downarrow}$ and $\ket{\uparrow}$ caused by the $A_\perp$ hyperfine term (see App.D). 

\begin{figure*}[t]
    \includegraphics[width=0.9\textwidth]{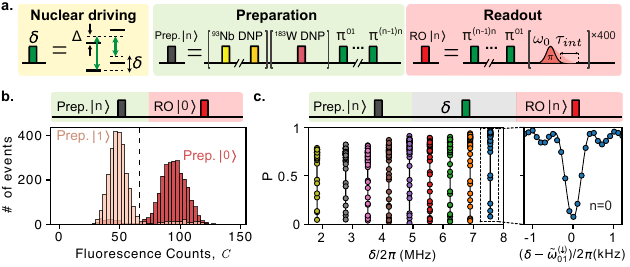}
    \caption{\label{fig2}
    \textbf{\Nb nuclear spin control, preparation, and readout.}
    \textbf{a.} Pulse sequences for \Nb nuclear spin driving, preparation and readout. 
    \textit{(left)} Coherent driving of the NMR transitions is achieved by microwave stimulated Raman driving while the \Er is in its ground state $\ket{\downarrow}$. To drive the $\ket{n} \leftrightarrow \ket{n+1}$ transition, we set the two-photon detuning $\delta$ to be on resonance with the corresponding transition, and we set the Raman detuning $\Delta = \omega_{01}^{(\uparrow)}/2$. Square pulses with sine-shaped rise and fall are used~\cite{osullivan_individual_2024}.
    \textit{(middle)} State preparation in $\ket{\downarrow,0}$ is achieved by driving all double-quantum transitions of the \Er-\Nb coupled system with chirped pulses of 1 ms duration and 100 kHz chirp range near resonance, with a waiting time of 1 ms between successive chirped pulses.(see App.D). The \W nuclear spin bath is then polarized (see App.F). Finally, a series of $\pi$ pulses transfers the nuclear spin population from $\ket{0}$ to $\ket{n}$.
    \textit{(right)} Readout of state $\ket{n}$ is achieved by first transfering the population from $\ket{n}$ to $\ket{0}$ using a sequence of $\pi$ pulses, followed by application of $400$ readout pulses. Each readout pulse is a $\pi$ pulse at $\omega_0$ with a duration of 40 \textmu s, and is followed by a delay of 1 ms to allow for \Er spin relaxation. The total number of counts $C$ is recorded.
    \textbf{b.} Single-shot readout of the \Nb spin state. 
    \textit{(top)} Pulse sequence for \Nb readout histogram. \Nb is prepared either in $\ket{0}$ or $\ket{1}$, before reading out $\ket{0}$. 
    \textit{(bottom)} The histogram of $C$ is shown for preparation in $\ket{0}$ (red) and in $\ket{1}$ (pink). A vertical dashed line marks the threshold used for state assignment.
    \textbf{c.} 
    \textit{(top)} Pulse sequence for $\ket{\downarrow,n} \leftrightarrow \ket{\downarrow,n+1}$ spectroscopy. \Nb is first prepared in state $\ket{n}$. A Raman pulse of duration 1 ms is then applied, and the two-photon detuning $\delta$ is varied. Finally the \Nb population in state $\ket{n}$ is read out.
    \textit{(bottom)} Microwave stimulated Raman spectroscopy of all NMR transitions for the \Er in $\ket{\downarrow}$. The inset shows the $01$ spectrum. The solid black line is a cardinal sine fit of the data, yielding the ac-Zeeman-shifted $01$ frequency $\tilde{\omega}_{01}^{(\downarrow)}$.
    }
\end{figure*}

We now describe nuclear spin driving, preparation in a given state $\ket{n}$, and readout, of the spin-9/2 impurity (see Fig.\ref{fig2}a). Coherent driving of the NMR transitions is achieved by stimulated Raman driving at microwave frequency~\cite{osullivan_individual_2024}. State preparation relies on dynamical nuclear polarization by the solid-effect, using the double- and zero-quantum transitions. Their frequency is first determined by spectroscopy (see App.\ref{app:zero-double-quantum-transitions}). State preparation in $\ket{0}$ is achieved by driving all double-quantum transitions of the spin-9/2 impurity with chirped pulses. Then, the $^{183}\mathrm{W}$ nuclear spin bath is also polarized, for linewidth narrowing (see App.\ref{app:dnp} and \cite{travesedo_all-microwave_2024}). Finally, a series of $\pi$ pulses transfers the nuclear spin population from $\ket{0}$ to $\ket{n}$. Readout on state $\ket{n}$ is achieved by first transfering the population from $\ket{n}$ to $\ket{0}$ using a sequence of $\pi$ pulses, followed by application of $400$ successive $\pi$ pulses at $\omega_0$ and counting the number of clicks, $C$. Owing to the long lifetime of state $\ket{0}$, which minimizes cross-relaxation, well-contrasted histograms are observed (see  Fig.~\ref{fig2}b), indicating good preparation and readout fidelity in $\ket{0}$.

The spectrum of all ground state manifold NMR transitions is shown in Fig.\ref{fig2}c. Their frequencies span a large range, from $1.5$\,MHz to $8$\,MHz, indicating that the quadrupolar interaction has comparable magnitude to the Zeeman energy. The $\ket{\downarrow,4} \leftrightarrow \ket{\downarrow,5}$ transition frequency is particularly interesting as it is first-order insensitive to the quadrupolar interaction, and its frequency is therefore close to the Zeeman frequency $\omega_I + A_\parallel/2$. By comparison with the gyromagnetic ratio of stable nuclear spins $9/2$, we determine that the impurity is a $^{93}\mathrm{Nb}$ atom (see App.\ref{app:nb_id}). 
This impurity may originate from the crystal growth, or from the niobium film sputtering during resonator fabrication. We hypothesize that the niobium atom enters in substitution of a $\mathrm{W}^{6+}$ atom in a pentavalent state, thus compensating the \Er extra positive charge compared to $\mathrm{Ca}^{2+}$. From $A_\parallel/2\pi \sim 130$\,kHz (see Fig.\ref{fig1}c), and $A_\perp/2\pi \sim 55$\,kHz (estimated from the sideband Rabi frequency, see App.\ref{app:zero-double-quantum-transitions}), we determine that the \Nb is located 0.57\,nm from the \Er along the $c$ axis, as shown in Fig.\ref{fig1}b. We confirm this assignment by measuring $A_\parallel$ and $A_\perp$ for various angles $\theta$ (see App.\ref{app:angular_dependence}). 
The measured NMR frequency spectrum sheds light on the vastly different cross-relaxation lifetimes observed in Fig.\ref{fig1}c, since these lifetimes scale like $\sim \omega_{n,n+1}^{2}$, explaining why low-$n$ states appear more stable than higher-$n$ states (see App.\ref{app:zero-double-quantum-transitions}).

\begin{figure}[t!]
    \includegraphics[width=\columnwidth]{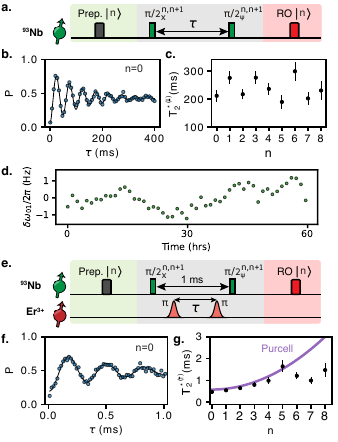}
    \caption{\label{fig3}
    \textbf{\Nb spin spectroscopy}.
    \textbf{a.} Ramsey pulse sequence between states $\ket{\downarrow,n}$ and $\ket{\downarrow,n+1}$. The \Nb spin is first prepared in $\ket{n}$. A $\pi/2$ of duration of 0.5 ms is applied, followed by an interpulse delay $\tau$.  The second $\pi/2$ pulse is applied with a linearly-increasing relative phase $\psi(\tau) = \omega_{IF} \tau$, with $\omega_{IF}$ chosen to avoid under-sampling
    \textbf{b.} Ramsey measurement with the \Er in $\ket{\downarrow}$. The probability to find the \Nb spin in $\ket{0}$ is shown as a function of $\tau$ for $n=0$. The solid black line is a cosine fit with a Gaussian decaying envelope yielding $T_2^{*(\downarrow)}=200$~ms and $\omega_{01}^{(\downarrow)} / 2\pi = 7560562.0(2)$ Hz.
    \textbf{c.} Coherence time $T_2^{*(\downarrow)}$ for each transition $\ket{\downarrow,n}\leftrightarrow\ket{\downarrow, n+1}$. Error bars correspond to the 1-$\sigma$ fit uncertainty. 
    \textbf{d.} Measured $\omega_{01}^{(\downarrow)}$ over 60 hours. The frequency drifts by ±1 Hz, indicating the spectral resolution of the ground-state measurements is limited by long-term magnetic drift.
    \textbf{e.} Excited-state frequency measurement sequence. The system is prepared in $|\downarrow,n\rangle$. Two $\pi/2$ pulses are applied on the $\ket{\downarrow,n}\leftrightarrow\ket{\downarrow, n+1}$ transition, separated by a fixed delay of $1$\,ms. Two non-selective microwave $\pi$ pulses of duration 5 $\mathrm{\mu}$s excite the \Er spin transiently into $\ket{\uparrow}$ during a varying time $\tau$.
    \textbf{f.} The probability to find the \Nb spin in $\ket{0}$ is shown as a function of $\tau$ for $n=0$. The solid black line is a cosine fit with a decaying exponential envelope yielding $T_2^{*(\uparrow)}=0.5$~ms and $(\omega_{01}^{(\uparrow)} - \omega_{01}^{(\downarrow)}) / 2\pi = -136547(45)$ Hz. 
    \textbf{g.} Coherence time $T_2^{*(\uparrow)}$ for each transition $\ket{\uparrow n}\leftrightarrow\ket{\uparrow, n+1}$. The purple solid line shows the dependence of the \Er relaxation time on the n-th allowed transition expected from the Purcell effect. Error bars correspond to the 1-$\sigma$ fit uncertainty. 
    }
\end{figure}

Precision measurement of the NMR spectrum requires pulsed spectroscopy using the Ramsey sequence, to avoid ac-Zeeman shifts caused by the application of the Raman drives~\cite{osullivan_individual_2024}. We first measure the ground-state-manifold frequencies $\omega^{(\downarrow)}_{n,n+1}$. We prepare the system in $\ket{\downarrow,n}$, apply two Raman $\pi/2$ pulses on the $\ket{\downarrow,n} \leftrightarrow \ket{\downarrow,n+1}$ transition separated by a time $\tau$, and measure the resulting probability to find the \Nb in $\ket{n}$ (see Fig.\ref{fig3}a.). For increased spectral resolution, we moreover polarize the $^{183}\mathrm{W}$ nuclear spin bath by solid-effect DNP at the beginning of each sequence (see Fig.~\ref{fig2} and App. F).
All curves are measured in an interleaved manner, in order to avoid possible drifts. Data are shown in Fig.\ref{fig3}b. for $n=0$, together with an exponential fit. We find coherence times $T_2^{*(\downarrow)} $ ranging between 200 and 300\,ms on all the transitions, corresponding to a standard deviation on the inferred frequency of $\pm 1$\,Hz (see Fig.\ref{fig3}c.). This indicates that magnetic noise is the dominant contribution to the Ramsey dephasing, since electric noise would impact more strongly transitions with extremal values of $n$, as was observed in an individual $^{123}\mathrm{Sb}$ nuclear spin in silicon \cite{fernandez_de_fuentes_navigating_2024}. 

The excited-state manifold frequencies $\omega^{(\uparrow)}_{n,n+1}$ are then measured. The sequence includes two $\pi$ pulses on the \Er separated by $\tau$, applied in-between two Raman $\pi/2$ pulses kept at a constant time delay of $1$\,ms (Fig.\ref{fig3}e.). The oscillation frequency directly yields $\omega^{(\uparrow)}_{n,n+1}-\omega^{(\downarrow)}_{n,n+1}$ (see Fig.\ref{fig3}f. for $n=0$). The coherence time is limited by the \Er relaxation time, leading to a $\sim 30$\,Hz uncertainty on the frequency measurement. We observe that $T_2^{*(\uparrow)}$ slightly increases with $n$, due to the increased detuning of the $\ket{\downarrow,n} \leftrightarrow \ket{\uparrow,n}$ transition from the resonator (see Fig.\ref{fig3}g.). The increase is less than expected from the Purcell effect, suggesting that this \Er has a relatively short non-radiative lifetime, of order $\sim 3$\,ms, possibly linked to the proximal \Nb impurity.

\begin{figure*}[t]
    \includegraphics[width=\textwidth]{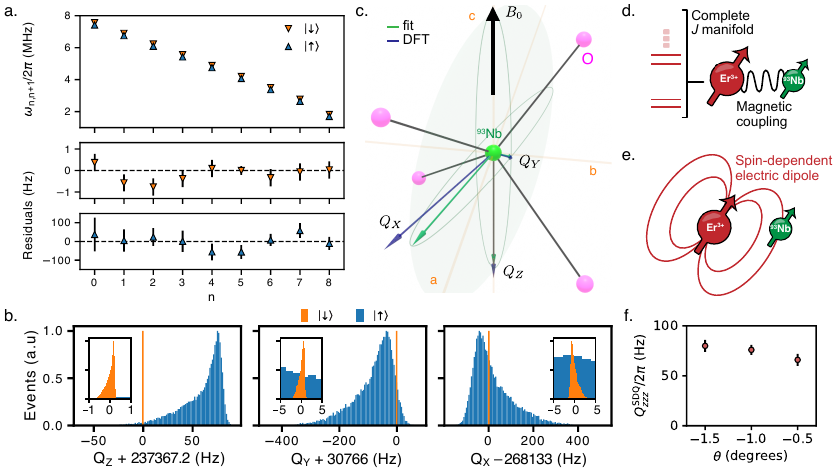}
    \caption{\label{fig4}
    \textbf{Nuclear transition frequencies and quadrupole fit}.
    \textbf{a.}
    \textit{(top)} Measured and fitted \Nb nuclear spin transition frequencies $\omega^{(\downarrow)}_{n, \ n+1}$ and $\omega^{(\uparrow)}_{n, \ n+1}$ (orange and blue triangles, resp.) as a function of $n$.
    \textit{(middle)} Orange triangles are residuals for the $\mathcal{H}^{(\downarrow)}$ fit. Error-bars are 1~Hz. Dashed line shows 0.
    \textit{(bottom)} Blue triangles are residuals for the $\mathcal{H}^{(\uparrow)}$ fit. Error-bars are 30~Hz. Dashed line shows 0.
    \textbf{b.} Posterior distribution of the fitted quadrupole tensor principal components $Q_{X,Y,Z}$ of $\mathcal{H}^{(\downarrow)}$ and $\mathcal{H}^{(\uparrow)}$ (orange and blue respectively). The distributions are normalized to the same height for visual clarity. Insets show zooms around $0$.
    \textbf{c.} Three-dimensional representation of the fitted (green) and computed (blue) quadrupole tensor. The \Nb nuclear spin (green) lies in the center, surrounded by the neighboring oxygen atoms (pink). The crystalline axes $a, b, c$ are shown as orange lines. 
    \textbf{d.} Schematics illustrating how magnetic coupling of the \Nb spin to higher-excited states of the \Er could lead to an effective spin-dependent quadrupole.
    \textbf{e.} Schematics illustrating how the electrostatic interaction between \Er and \Nb could lead to a spin-dependent quadrupole.
    \textbf{f.} Fitted (red circles) spin-dependent quadrupole $Q_\mathrm{sdq}$ as a function of $\theta$.
    }
\end{figure*}

We now analyze the frequency data to extract the nuclear spin parameters for each \Er spin state, according to Hamiltonian Eq.~\ref{eq2}. 
The quadrupolar tensor, being traceless and symmetric, comprises five independent variables. Our data are measured at only one magnetic field orientation, and we are therefore incapable of determining the overall rotation angle around the nuclear spin quantization axis, leaving four quadrupolar parameters to determine in addition to the nuclear Zeeman frequency (see App.\ref{app:spin_hamiltonian}). These $5$ parameters are obtained for each \Er spin orientation, by diagonalizing the Hamiltonians $\mathcal{H}^{(\downarrow)}$ ($\mathcal{H}^{(\uparrow)}$) and minimizing the difference between the computed and measured frequencies of the $9$ transitions in each manifold. The fit is performed through a Monte-Carlo Markov-Chain (MCMC) evolution, and the uncertainty of the fitted parameters is computed from the posterior distributions. More details about the fitting procedure can be found in App.\ref{app:quadrupole_fitting_procedure}.


The fitted and measured frequencies are shown in Fig.\ref{fig4}a. They agree up to statistical uncertainty ($1$\,Hz for the $\downarrow$ data, $30$\,Hz for the $\uparrow$ data). 
The fit returns the quadrupolar tensors $\bar{\bar{Q}}^{(\downarrow)}$ and $\bar{\bar{Q}}^{(\uparrow)}$, up to a rotation around the $z^{(\downarrow)}$ and $z^{(\uparrow)}$ axes respectively. Diagonalization of these tensors reveals that the principal axis with the largest eigenvalue (denoted as $X$) is approximately located within the crystal's $(a,b)$-plane. The second-strongest principal axis (denoted as $Z$) is approximately parallel to the crystal $c$-axis, and therefore also to $B_0$. 
The posterior distribution of the quadrupole principal values are shown in Fig.\ref{fig4}b for both \Er spin states. We observe a statistically-significant ($3\sigma$) difference between the values of $Q_{Z}^{(\uparrow)}$ and $Q_{Z}^{(\downarrow)}$. Due to the larger statistical uncertainty on $Q_{X}$ and $Q_{Y}$, this observation is not reproduced on $X$ or $Y$.

Our results therefore indicate the existence of a new term in the spin Hamiltonian Eq.~\ref{eq:SpinHam}, which modifies the nuclear quadrupolar interaction depending on the electron spin state. Given that our measurements are maximally sensitive along the $z$ axis, we model this interaction with Hamiltonian 

\begin{equation}\label{eq:SDQ}
\mathcal{H}_\mathrm{sdq} = Q_\mathrm{sdq}\, S_z \tfrac12 [3 I_z^2 - I(I+1) ].
\end{equation}



To further test our model, we take a step back in the approximations and fit the complete frequency dataset with the following Hamiltonian

\begin{align}
\begin{split}
        \mathcal{H}/\hbar &= \omega_S\cdot S_z + \omega_I\cdot I_z +\mathbf{S} \cdot \bar{\bar{A}} \cdot \mathbf{I} \,+ \\ 
         &+ \mathbf{I} \cdot \bar{\bar{Q}} \cdot \mathbf{I} + \mathcal{H}_\mathrm{sdq}  + \mathcal{H}_\mathrm{L}.
\label{eq:Hcomplete}
\end{split}
\end{align}

\noindent Here, the full hyperfine tensor $\bar{\bar{A}}$ is considered. $A_\parallel$ is kept as a fit parameter; $A_\perp/2\pi=55(9)$~kHz is inferred from measurements; the non-secular terms are calculated considering the magnetic dipolar interaction in the point-dipole approximation (see App.\ref{app:spin_hamiltonian}). We also take into account small frequency shifts of the levels $\ket{\uparrow,n}$ caused by the coupling to the vacuum fluctuations in the resonator, via the Lamb shift Hamiltonian $\mathcal{H}_{\text{L}} = \Delta \omega_n \ket{\uparrow,n}\bra{\uparrow,n}$, with $\Delta \omega_n = g_0^2 \frac{\Delta_n}{\Delta_n^2 + \kappa^2/4}$ and  $\Delta_n = \omega_0 - \omega_n$ (see App \ref{app:spin_hamiltonian}). Note that because the Hamiltonian Eq.~\ref{eq:Hcomplete} is not invariant by rotation around the $z$ axis (contrary to Eqs.~\ref{eq2}), the full quadrupolar tensor can now be obtained. We fit $8$ parameters from Hamiltonian Eq.~\ref{eq:Hcomplete} ($\omega_I$, $A_{\parallel}$, the $5$ quadrupolar parameters, and $Q_\mathrm{sdq}$) to the $18$ measured frequencies (see App.\ref{app:quadrupole_fitting_procedure}). In the fitting process, $A_\perp$ is sampled from a Gaussian distribution to take into account its uncertain value. We obtain $\omega_I/2\pi =- 4732.618(4)$~kHz, $A_{\parallel}/2\pi=133.497(8)$~kHz, $C_q / 2\pi = 19.30470(6)\ \mathrm{MHz}$, and $\eta = 0.770387(6)$. It is worthwhile noting that in pure \Ca, the quadrupolar tensor of a \Nb impurity replacing a \W would be axially symmetric around the $c$ axis due to the $S_4$ symmetry of these sites ($\eta = 0$). The strong biaxiality measured (with $\eta \sim 0.8$) is therefore a signature of the $S_4$ symmetry breaking, caused by the replacement of a calcium by an erbium atom in close vicinity to the \Nb.

The fitted quadrupolar tensor $\bar{\bar{Q}}$ is graphically represented in the crystal frame in Fig.\ref{fig4}c. We also compare it to DFT simulations on a system similar to ours, with the \Er replaced by a $\mathrm{Y}^{3+}$ for easier calculation (see App.\ref{app:DFT_calculation}). They yield $C_{q,\text{DFT}}/2\pi=22.7$~MHz, $\eta_{\text{DFT}}=0.76$, and similar orientation of the principal axes, in approximate agreement with the values inferred from measurements. Finally, the SDQ term $Q_\mathrm{sdq}/2\pi= 66(6)$~Hz is fitted to a non-zero value with a high degree of confidence. We check the consistency and reproducibility of our analysis by repeating these measurements for a small range of in-plane orientation $\theta$ of the magnetic field. Despite the vastly different values of the Hamiltonian parameters (in particular, $\omega_I$ and $A$), the values extracted for $Q_\mathrm{sdq}$ remain approximately unchanged. 


We identify two possible mechanisms that could account for the SDQ term. The first one is the magnetic coupling of the \Nb to higher-excited states of the \Er (see Fig.~\ref{fig4}d). Such coupling is well-known to cause nuclear frequency shifts that mimic a quadrupolar interaction, leading to a so-called pseudo-quadrupolar contribution of magnetic origin \cite{foley_second-order_1947, abragam_electron_2012}. This pseudo-quadrupolar component might itself depend on the \Er spin state, as envisioned in \cite{ghatikar_zeeman-field-dependent_1966}. To investigate the magnitude of this effect, we model the hyperfine interaction of the \Nb with the complete $J=15/2$ \Er manifold, using the crystal-field Hamiltonian of \Er:\Ca determined in Ref.~\cite{enrique_optical_1971} (see App.\ref{app:pseudo_quadrupole_interaction}). 
We predict a pseudo-SDQ of $ -10$\,Hz, an order of magnitude smaller than the data and with opposite sign, indicating that this mechanism should be considered but cannot be the only one at play. 

The electrostatic interaction between the \Er and the \Nb spins is another mechanism that may contribute to the SDQ term (see Fig.~\ref{fig4}e). Bloembergen\ \cite{noauthor_national_1961,Royce.PhysRev.131.1912} predicted that paramagnetic centers located at sites without inversion symmetry carry an electric dipole whose magnitude is proportional to $B_0$ and whose orientation changes with the spin state. This leads to linear electric-field shifts of the EPR transition frequency that have been observed in numerous paramagnetic systems~\cite{ludwig_splitting_1961}, including in \Er:\Ca \cite{mims_electric_1964, mims_electric_1965}. The \Er spin states therefore also possess a small permanent electrical dipole, of magnitude proportional to $B_0$ and orientation dependent on the spin state. This dipole produces a spin-dependent electric-field at the \Nb site. Its gradient is several orders of magnitude too small to account for the measured value of $Q_\mathrm{sdq}$ (see App.\ref{app:electric-dipole_calculation}); however, the electric field itself may induce a quadrupole shift since the \Nb site is also not inversion-symmetric~\cite{noauthor_national_1961,armstrong_linear_1961,asaad_coherent_2020}, leading to the SDQ effect. We estimate the magnitude of the electric field generated by the \Er at the \Nb location to be $\sim 100\,\mathrm{V/cm}$ (see App \ref{app:electric-dipole_calculation}). Although the electric quadrupole sensitivity of \Nb:\Ca is not known, values ranging between $0.1$ and $1$ Hz/(V/cm) were reported for a variety of nuclei \cite{macfarlane_optical_2014}. Applied to \Nb, we get an estimated SDQ comparable to the measured value. Our data therefore suggest the existence of a new type of coupling between a paramagnetic center and a nuclear spin, of electrostatic nature. 


We now discuss the generality of our findings. The spin-dependent pseudo-quadrupole mechanism should be sizeable in paramagnetic centers with low-lying excited states, such as metallic centers and particularly lanthanides. The direct electrical coupling mechanism requires paramagnetic centers with spin-orbit coupling and non-inversion symmetry, also found in metallic centers. Overall, measurable SDQ terms should therefore be found mainly in systems involving metallic paramagnetic centers (molecules or impurities in solids), but not in organic radicals. Larger SDQ values than the one reported here should be observable on the nuclear spin belonging to the metallic center itself, for instance in $^{167}\mathrm{Er}^{3+}$.

\begin{figure*}[t]
    \includegraphics[width=\textwidth]{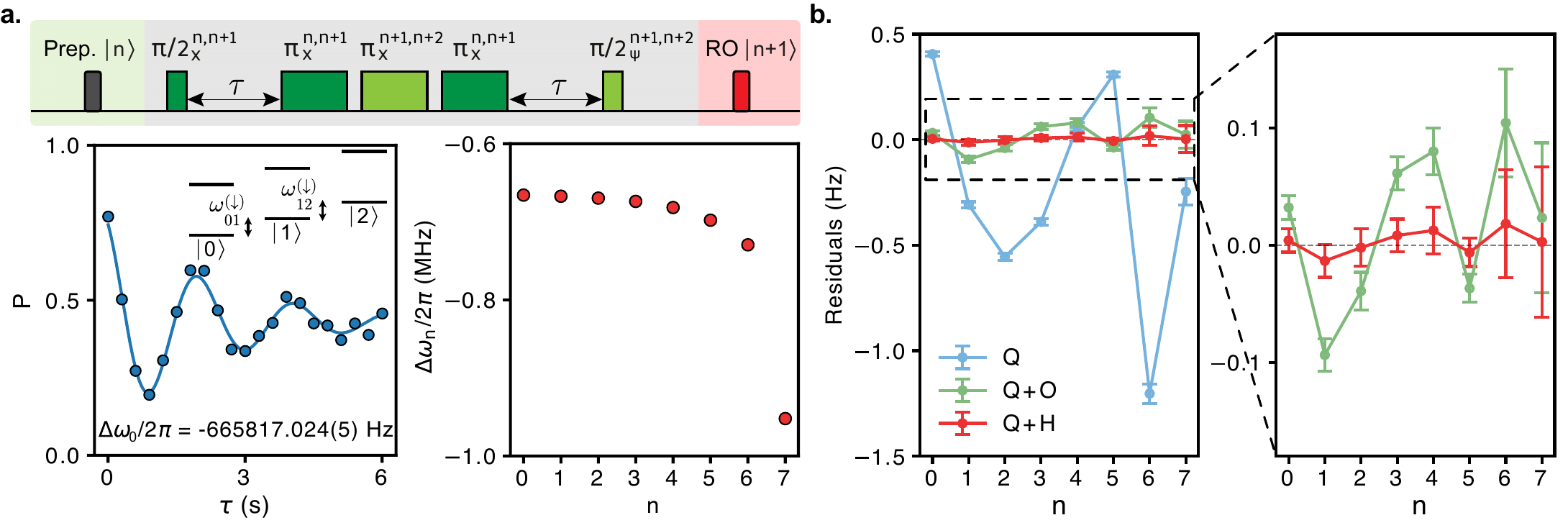}
    \caption{\label{fig5} \textbf{Correlated echo and multipole fit.}   
    \textbf{a.}
    \textit{(top)} Correlated-echo pulse sequence. After preparation in $|\downarrow,n\rangle$, a $\pi/2$ is applied on $|\downarrow,n\rangle \leftrightarrow |\downarrow,n+1\rangle$. After a waiting time $\tau$, a sequence of $3$ $\pi$ pulses transfer the coherence to the $n+1,n+2$ transition. A $\pi/2$ pulse is applied after a delay $\tau$ on $|\downarrow,n+1\rangle \leftrightarrow |\downarrow,n+2\rangle$ with a phase $\psi = \omega_{IF}\tau$, $\omega_{IF}$ being chosen to avoid under-sampling. 
    \textit{(left)} Correlated-echo data on the $0\leftrightarrow1,1\leftrightarrow2$ transition (see level diagram). Blue circles are measured probabilies to find the \Nb spin in $\ket{1}$ as a function of $\tau$. Solid blue line is an exponentially-decaying cosine fit, yielding $\Delta \omega_0 = -665817.024(5)$\,Hz.
    \textit{(right)} The red circules show the measured $\Delta \omega_n$ as a function of n. The error-bars are not visible on that scale.
    \textbf{b.} Residuals of the Hamiltonian fit (see text) with three different models: quadrupole (blue), quadrupole with octupole (green), and quadrupole with hexadecapole (red). The error-bars are measurement uncertainties obtained by bootstrapping. Right panel is a zoom. 
    }
\end{figure*}

The spectral resolution of the ground-state measurements is limited by long-term magnetic drift, which manifests itself by a slow change of the measured NMR frequencies by $\pm 1$\,Hz over a scale of several tens of hours (see Fig.\ref{fig3}d. for data on the $0\leftrightarrow 1$ transition). Since the magnetic field sensitivity of neighboring transitions ($n\leftrightarrow n+1$ and $n+1\leftrightarrow n+2$) is approximately the same, it is possible to measure the difference between these frequencies with a spectral resolution higher than can be reached for the frequencies themselves. In that purpose, we use the pulse sequence shown in Fig.\ref{fig5}a, inspired by refs.~\cite{degen_creation_2021, liao_nuclear_1994,abobeih_atomic-scale_2019}. It consists in a Hahn echo sequence in which the phase accumulates on the $n\leftrightarrow n+1$ transition during the first half, and on the $n+1\leftrightarrow n+2$ in the second half. The resulting echo phase oscillates as $(\omega_{n+1,n+2}^{(\downarrow)} - \omega_{n,n+1}^{(\downarrow)}) \tau$, $\tau$ being the half-echo duration, as seen in Fig.\ref{fig5}a. for the $0\leftrightarrow1,1\leftrightarrow2$ transitions. Because magnetic drift is largely mitigated, the signal decays in $3$\,s, much longer than $T_2^*$. This corresponds to a resonance line of FWHM $0.1$\,Hz whose center can be determined with $10$\,mHz resolution, given the experimental signal-to-noise ratio.

The frequency difference between neighboring transitions $\Delta \omega_n \equiv \omega_{n+1,n+2}^{(\downarrow)} - \omega_{n,n+1}^{(\downarrow)} $ can then be fitted to the simple spin Hamiltonian model Eq.~\ref{eq:Hcomplete}, using the $4$ quadrupole tensor parameters as adjustable variables. The residuals, shown in Fig.\ref{fig5}b, indicate that the model fails to reproduce the data in a statistically-significant manner, and that higher-order terms are needed. An octupolar (O) contribution could exist, due to the proximity of the \Nb to the \Er paramagnetic impurity. The latter may produce a sizeable magnetic field second derivative at the \Nb location, which may couple to a possible \Nb magnetic octupolar moment. A hexadecapolar (H) contribution could also exist, due to the coupling of the crystalline electric field third derivative to a possible \Nb electric hexadecapolar moment. We test the two models separately, by including in the fit either the  term $\tfrac{C_3}{I(2I-1)(I-1)}  I_z^3$ (model $O$), or the term $\tfrac{C_4}{I(2I-1)(I-1)(2 I -3)}  I_z^4$ (model $H$). While the $O$ model with $C_3/2\pi = 24.1(1)$\,Hz improves the residuals, a statistically-significant discrepancy remains. On the other hand, the $H$ term with $C_4/2\pi = 9.6(1)$\,Hz yields good agreement with the data. To the best of our knowledge, this constitutes the first observation of a nuclear hexadecapole interaction in the solid-state~\cite{segel_nuclear_1978,doering_search_1986,liao_nuclear_1994}. 

While this interaction term could originate from the coupling of the hexadecapolar moment of the \Nb nucleus to the 3rd derivative of the electric field, it could also arise from the coupling to one or several higher-energy states. Such pseudo-hexadecapolar interaction was invoked for $\mathrm{Li I}$ molecules measured in a molecular-beam~\cite{cederberg_evidence_1999,thyssen_quadrupole_2001}. In our case, the hyperfine coupling to the \Er excited state is obviously relevant, and we have therefore estimated its contribution by numerical diagonalization of the complete spin Hamiltonian (see App.\ref{app:pseudo_quadrupole_interaction} for more details). We find the pseudo-hexadecapole to be of $-0.03$\,Hz, implying that it cannot account for our measurements, since this value is more than two order of magnitude smaller and with the opposite sign. Further work will be needed to determine whether the value measured here is compatible with the expected magnitude of a hexadecapolar moment in \Nb. More generally, these results establish our measurement method as a viable route to study hexadecapolar contributions in a variety of nuclei in the solid-state.

The detection of single paramagnetic centers by microwave photon counting only requires sufficiently long spin-lattice relaxation times at 10mK, and should therefore be applicable to a large variety of systems and samples~\cite{wang_single-electron_2023}. Our demonstrated ability to characterize chemically and structurally an unknown nuclear-spin-carrying atomic impurity in the vicinity of a paramagnetic center opens interesting perspectives, such as the ability to characterize the constituent atoms of individual molecules for instance. The high spectral resolution reached in the \Nb levels spectroscopy should be achievable more generally in arbitrary nuclear spin systems, as long as they can be coupled to a paramagnetic center. This would be particularly relevant for quadrupolar nuclei in disordered systems (molecules in frozen solutions for instance), where high spectral resolution is difficult to achieve by conventional ensemble methods. Our observation of a spin-dependent quadrupole calls for further theoretical work in order to confirm the electrical coupling mechanism. The effect should also be studied in other coupled spin systems. Finally, the observation of a nuclear hexadecapole interaction in the solid-state opens the way to a more systematic study of this interaction in a variety of nuclei, which could bring new insight into the understanding of the shape of deformed nuclei. 

\section*{Acknowledgements}
{We acknowledge technical support from P.~Simon, and are grateful for fruitful discussions within the Quantronics group, as well as with T. Taminiau who suggested the correlated echo sequence. We acknowledge support of the R\'egion Ile-de-France through the DIM QUANTIP, from the AIDAS virtual joint laboratory, from the France 2030 plan under the ROBUSTSUPERQ (ANR-22-PETQ-0003), NISQ2LSQ (ANR-22- PETQ-0006), and QMEMO (ANR-22-PETQ-0010) grants. This project has received funding from the European Union Horizon 2020 research and innovation program under the project OpenSuperQ100+ and from the European Research Council under the grant no. 101042315 (INGENIOUS). We thank the support of the CNRS research infrastructure INFRANALYTICS (FR 2054) and Initiative d'Excellence d’Aix-Marseille Université – A*MIDEX (AMX-22-RE-AB-199). We acknowledge IARPA and Lincoln Labs for providing the Josephson Traveling-Wave Parametric Amplifier. We acknowledge the crystal lattice visualization tool VESTA.}

\section*{Author contributions}
{The experiment was designed by J.T., Z.W.H., J.O'S., E.F., and P.B. The crystal was grown by P.G. and characterized by EPR spectroscopy by S.B. The spin resonator chip was designed and fabricated by J.T. with the help of P.A. The SMPD was designed, fabricated, and characterized by L.P. under supervision of E.F. Data were acquired by J.T. and Z.W.H., with the help of L.M. and N.T. Data analysis and simulations were conducted by J.T. and Z.W.H., with the help of L.M., under supervision of P.B. DFT calculations were performed by T. Charpentier. The crystal field simulation was written by T. Chaneliere. The manuscript was written by J.T., Z.W.H, and P.B., with contributions from all co-authors. The project was supervised by J.O'S., P.B. and E.F.}

\section*{Competing interests}
The authors declare no competing interests.

\clearpage
\begin{appendix}

\onecolumngrid

\section{Sample and experimental setup}
\label{app:sample_details}

\begin{figure}[b]
    \includegraphics[width=\columnwidth]{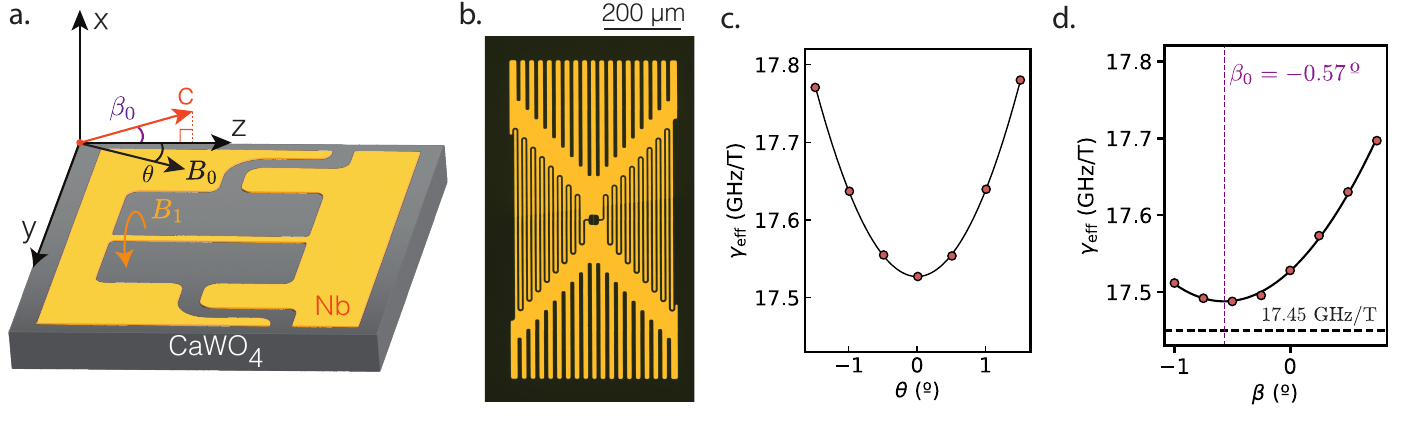}
    \caption{\label{sfig:resonator}
    \textbf{Sample schematic and false color micrograph}.
    \textbf{a.} Sample schematics. A resonator (yellow) is fabricated out of a niobium thin-film, on the surface of a CaWO$_4$ slab (grey). The resonator's plane has a out-of-plane angle $\beta_0$ with respect to the $(a,c)$-plane of the crystal. A magnetic field $B_0$ is applied in the plane of the resonator, with an angle $\theta$ with respect to the projection of the $c$-axis on the resonator plane. The nanowire constriction generates an oscillating magnetic field $B_1$.
    \textbf{b.} False color micrograph of an equivalent Nb thin-film resonator (yellow) fabricated on a CaWO$_4$ crystal (black). The nanowire lies in the center between the two plates, marked with a red line. The picture is stitched from 2 independent images.
    \textbf{c.} Measured effective gyromagnetic factor of the \Er:\Ca bulk resonance as a function of $\theta$. A parabolic fit (solid dashed line) is plot along the measurement data (red dots).
    \textbf{d.} Effective gyromagnetic factor of the \Er:\Ca bulk resonance as a function of $\beta$. The black dashed line is the literature value for the $c$-axis component of the \Er:\Ca gyromagnetic tensor \cite{antipin_aa_anisotropy_1981}. The purple dashed line marks the angle between the c-axis and the resonator plane $\beta_0 = -0.57^\circ$.
    }
\end{figure}

\paragraph{Sample} The sample was grown at Institut de Recherche de Chimie Paris, cut into a slab of dimensions 7x4x0.5~mm, and polished. The sample surface approximately corresponds to the $(a,c)$-plane of the crystal, with the shorter edge approximately parallel to the $c$-axis. The resonator is patterned out of a 50nm-thick niobium film deposited on the surface. Figures \ref{sfig:resonator}.a and \ref{sfig:resonator}.b depict a sample schematic as well as a false color micrograph respectively. More details about the resonator design and fabrication process can be found in refs. \cite{travesedo_all-microwave_2025} and \cite{osullivan_individual_2024} where the same sample was used.

\begin{figure}[t]
    \includegraphics[width=\textwidth]{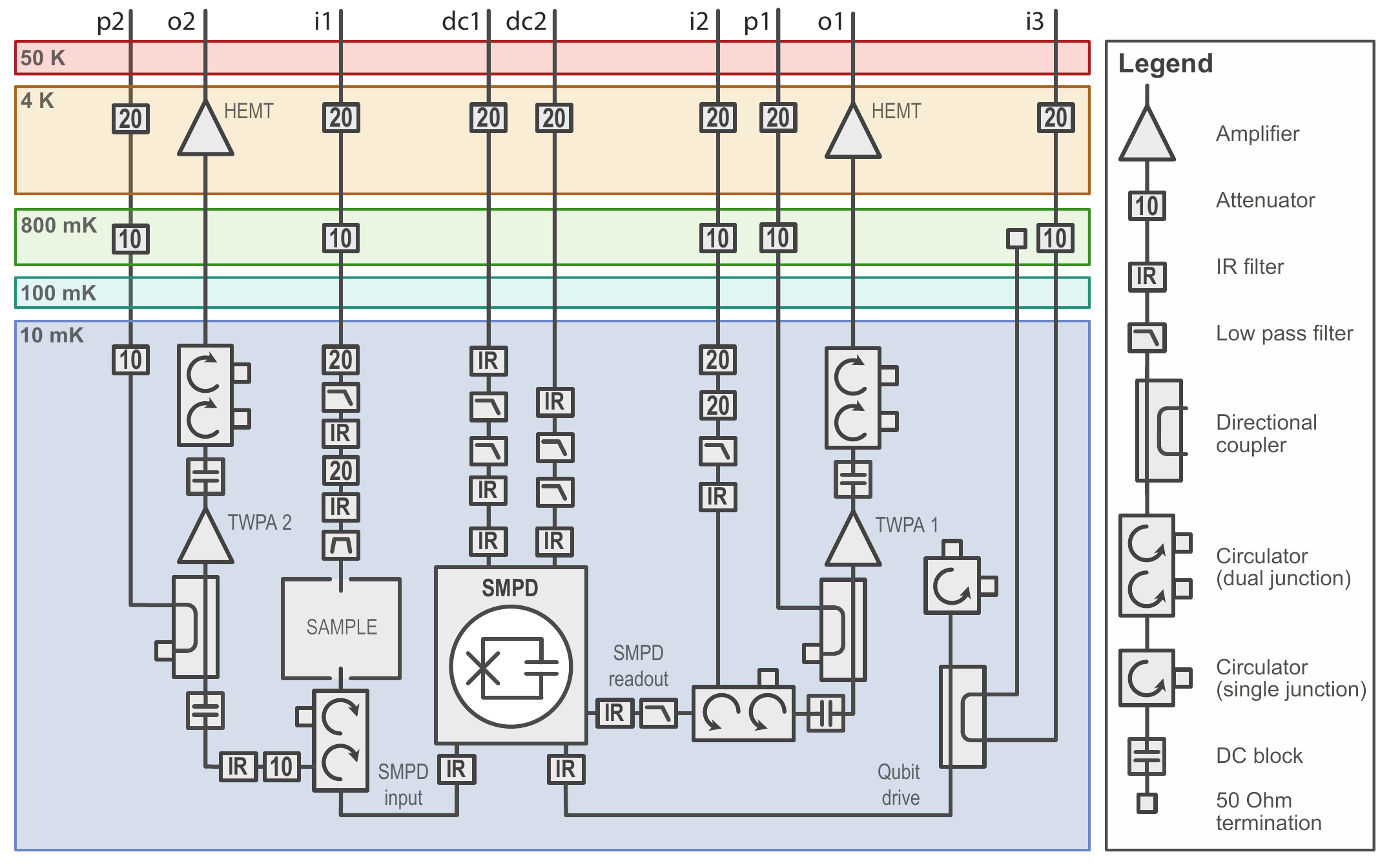}
    \caption{\label{sfig:wiring_diagram}
    \textbf{Setup}. Schematic of the cryogenic wiring. The sample is hosted inside of a 3D copper cavity, interfaced via two antennas in a transmission scheme.
    }
\end{figure}

\paragraph{Setup} A full diagram of the cryogenic microwave setup is shown in Fig.\ref{sfig:wiring_diagram}.

The sample is inserted into a copper sample holder and cooled down to 10~mK in a dilution refrigerator. The resonator is capacitively coupled to two antennas, with coupling rates dissimilar by a factor approx. $10$. 

Compared to Ref.~\cite{travesedo_all-microwave_2025}, the line ($i1$) used to transmit the microwave pulses driving the spins is now distinct from the SMPD line in the hope of minimizing dark counts induced by the application of microwave pulses. This line is  heavily attenuated at low temperatures to thermalize the microwave field at the lowest possible temperature. $i1$ is connected to the weakly-coupled input port, whereas the strongly-coupled port is directed towards the SMPD input. In this way, most photons emitted by the spins are directed towards the SMPD input. 

\paragraph{SMPD} The SMPD has been described in Ref. \cite{pallegoix_enhancing_2025}. The SMPD pump is sent through line $i3$. The frequency and bandwidth are tuned by sending current through lines $dc1$ and $dc2$, respectively. The SMPD input line being filtered by the resonator, we instead use the qubit drive line ($i3$), coupled to the buffer input by weak capacitive coupling, to tune the SMPD parameters (pump frequency and amplitude) for maximum detection efficiency. Readout pulses are sent through line $i2$, and the reflected signals are directed to output line $o1$, which contains a TWPA1 pumped through line $p1$. In this work, the SMPD bandwidth is set to 300~kHz resulting in an efficiency of $\simeq 0.79$ and a dark count rate of 40(5) counts per second. The duration of a detection cycle is approximately $\sim$17~\textmu s, with small variations appearing due to the active reset on the qubit. The readout and reset last 2~\textmu s, which corresponds to the detector dead-time. 

\paragraph{Resonator characterization} We first characterize the resonator frequency using a vector network analyzer (VNA). A microwave tone is injected through line $i1$, and the signal transmitted through the resonator is amplified by TWPA2, pumped via line $p2$. To tune the single-microwave-photon detector (SMPD) input frequency to $\omega_\mathrm{r}$, we subsequently fine-tune the SMPD frequency by sweeping a weak probe tone around $\omega_\mathrm{r}$ on $i1$ while varying the voltage applied to the DC tuning line $dc1$. The probe tone is filtered by the resonator before entering the SMPD, while the DC voltage tunes the SMPD transition frequency.

\begin{figure}[t]
    \includegraphics[width=\columnwidth]{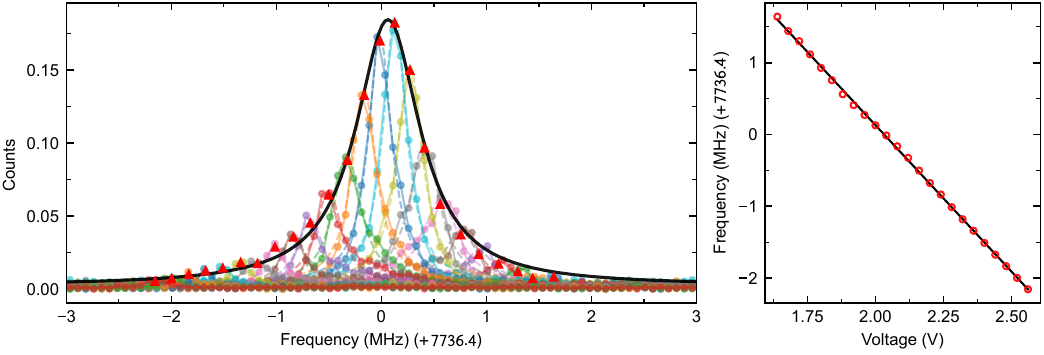}
    \caption{\label{sfig:buffer_autocalibration}
    \textbf{SMPD–resonator frequency calibration}.
    \textbf{a.} Response of the single-microwave-photon detector (SMPD) as a function of probe frequency for different DC tuning voltages (colored open circles). Each trace is fitted to a Lorentzian (dotted lines), yielding narrow linewidths of approximately 300 kHz. The extracted peak centers (red triangles) form an envelope reflecting the resonator response; fitting this envelope yields a resonator linewidth of 740 kHz. A slight asymmetry in the envelope is attributed to weak parasitic transmission and is well described by a Fano lineshape, with negligible impact on the extracted linewidth.
    \textbf{b.} SMPD resonance frequency as a function of DC tuning voltage, showing a linear dependence used to align the SMPD to the resonator frequency.
    }
\end{figure}

The SMPD response for different DC voltages appears as a set of narrow resonance lines with linewidths of approximately 300 kHz, shown in different colors in Fig.\ref{sfig:buffer_autocalibration}\textbf{a}. Each individual trace is well described by a Lorentzian lineshape, and the extracted peak positions (red triangles) form a broader envelope corresponding to the resonator response. Fitting this envelope yields a resonator linewidth of $\kappa/2\pi = 740$\,kHz. We observe a slight asymmetry in the envelope lineshape, which we attribute to weak parasitic transmission interfering with the resonator response. This asymmetry is well captured by a Fano model; however, the extracted linewidth agrees with the Lorentzian fit within uncertainty, and the large Fano asymmetry parameter indicates that the parasitic contribution is weak and does not affect the linewidth determination.

In Fig.\ref{sfig:buffer_autocalibration}b, the SMPD resonance frequency is plotted as a function of the applied DC voltage and fitted with a linear relation. Using this calibration, we set the DC voltage corresponding to the resonator frequency, thereby tuning the SMPD in resonance with the resonator. This information is also used to tune the SMPD frequency across different electron transition frequencies in Fig.\ref{fig1}c since they span wider than the SMPD bandwidth.

\paragraph{Field alignment and coil calibration} The magnetic field is applied in the resonator plane. The measurements and analysis depend sensitively on the angle made by the magnetic field with the crystalline $c$-axis, which is therefore relevant to determine. 

We measure the angle $\beta_0$ between the resonator plane and the crystalline $c$ axis following the procedure described in \cite{wang_month-long-lifetime_2024}. We first find the projection of the $c$-axis on the resonator plane by finding the angle at which the effective gyromagnetic ratio $\gamma_\mathrm{eff}$ of the bulk \Er EPR line is the lowest (see Fig.\ref{sfig:resonator}c). We then measure $\gamma_\mathrm{eff}$ as a function of the out-of-plane angle, $\beta$. The c-axis then corresponds to the minimum value of the effective gyromagnetic ratio, which is found at $\beta_0=-0.57^\circ$ (see Fig.\ref{sfig:resonator}d). This measurement also calibrates the field generated by our coil, by comparison to the expected value of $\gamma_\parallel/2\pi = 17.45$\,GHz/T. We find the applied field is $1.002$ times larger than the nominal applied field, possibly due to a room-temperature cylindrical mu-metal magnetic shield surrounding the cryostat. This correction is taken into account as a global factor in all the measurements, so that we estimate the relative uncertainty on the applied field to be lower than $10^{-3}$.

\section{Spin Hamiltonian}
\label{app:spin_hamiltonian}

\begin{figure}[t]
    \includegraphics[width=\textwidth]{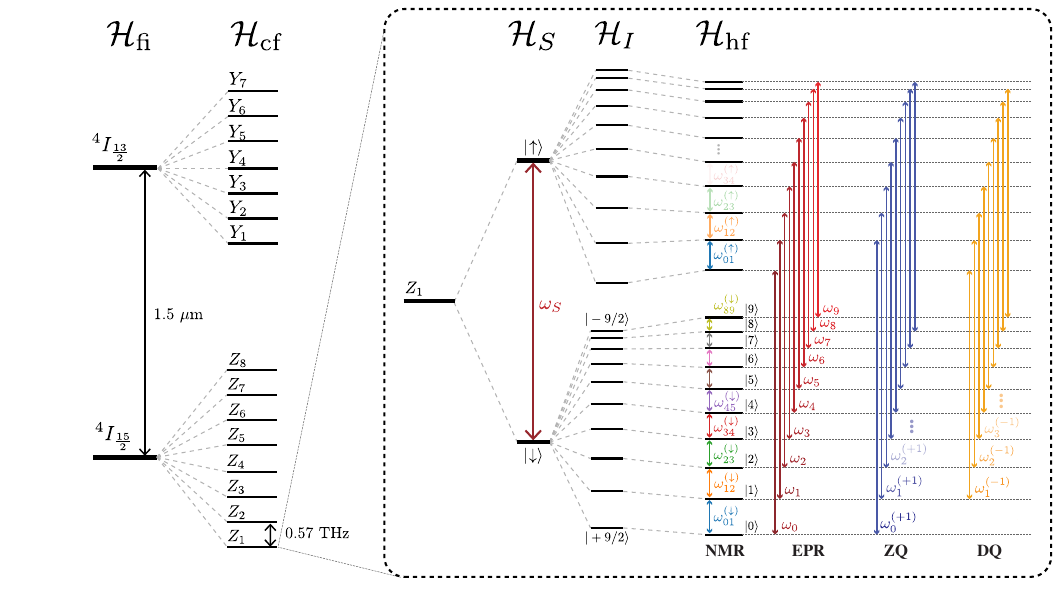}
    \caption{
    \textbf{Energy level structure}. Level splittings for every Hamiltonian term. From left to right, free-ion ($\mathcal{H}_{\text{fi}}$), crystal-field ($\mathcal{H}_{\text{cf}}$). Inset, effective spin-1/2 ($\mathcal{H}_{\text{S}}$), nuclear spin ($\mathcal{H}_{I}$) and hyperfine ($\mathcal{H}_{\text{hf}}$). NMR-allowed, EPR-allowed, zero-quantum (ZQ) and double-quantum (DQ) transitions are highlighted in the last energy level structure as multicolored, red, blue and orange arrows respectively.
    }
    \label{sfig:energy-levels}
\end{figure}

We discuss in detail the spin Hamiltonian of the \Nb -- \Er coupled system, and we discuss the approximations used in various places of the manuscript. We start with the \Er modeling, then the \Nb spin, and then their coupling. Note that the spin-dependent quadrupole is discussed in a later section (App.\ref{app:pseudo_quadrupole_interaction} \& \ref{app:electric-dipole_calculation}), and is not considered here. Fig.\ref{sfig:energy-levels} schematically depicts the various Hamiltonian contributions discussed below.

\subsection{Erbium}

\Er possesses $11$ valence electrons in the $4f$ shell. The energy eigenstates of the free-ion Hamiltonian $\mathcal{H}_{\mathrm{fi}}$ are degenerate manifolds of well-defined total angular momentum $J$. The $16$-fold-degenerate ground state manifold $J=15/2$ is separated by an optical transition at $1.5$~\textmu m from the lowest-excited manifold, $J=13/2$ (see Fig.\ref{sfig:energy-levels}).

In \Ca, \Er enters in substitution of $\mathrm{Ca}^{2+}$. The free-ion energy levels are perturbed by the electrostatic potential $V$ at the \Er location produced by the neighboring atoms of the lattice. This potential can be expressed as a multipole expansion, and written by separating terms that are even and odd under the inversion symmetry ($r \leftrightarrow -r$), $V = V_\text{odd} + V_\text{even}$. Because the \Er insertion site (of symmetry $S_4$) does not present inversion symmetry, $V_\text{odd} \neq 0$.

\paragraph{Odd-parity crystal-field terms and electric dipole} The effect of $V_\text{odd}$ has been discussed in \cite{judd_optical_1962,kiel_theory_1966}. This term couples electrons in the $4f$ shell with shells of opposite parity, for instance the $5d$. Because of the large energy difference, the wave-function admixture is small; nevertheless, its consequences are important. First, it causes a non-zero electric dipole matrix element between Kramers doublets (and in particular, it allows electric-dipole driving of the optical transition at $1.5$~\textmu m~\cite{judd_optical_1962}). Second, it implies that the eigenstates of \Er present a non-zero average electric dipole moment (which is relevant for this work). This moment is suppressed at zero magnetic field by Kramers degeneracy, but the latter is lifted under a finite magnetic field. Therefore, in finite magnetic fields, \Er eigenstates possess a non-zero electric dipole whose magnitude is approximately proportional to the magnetic field strength and whose direction is spin-dependent. We attribute part of the Spin-Dependent-Quadrupole effect to this electric dipole, as discussed in the main text and in App.\ref{app:electric-dipole_calculation}. 

\paragraph{Crystal field Hamiltonian} We now focus on the $4f$ part of the \Er electronic wavefunction, leaving aside the effect of the odd crystal-field terms. Non-zero matrix elements between $4f$ states require even parity operators; hence, the eigenstates wavefunctions and energy levels are determined by $V_\text{even}$. We moreover neglect the admixture induced by $V_\mathrm{even}$ between levels of well-defined angular momentum $J$, which is justified by the large energy difference between the $J$ manifolds. The effect of $V_\text{even}$ in the $J=15/2$ ground state is to split the 16-fold degenerate level into $8$ Kramers doublets (see Fig.\ref{sfig:energy-levels}).

Although the measurements reported in this work are performed at temperatures such that only the lowest-energy Kramers doublet ($Z_1$) is populated, the higher-energy doublets ($Z_{2..8}$) play a role in the pseudo-quadrupole effect (see App \ref{app:pseudo_quadrupole_interaction}). We describe these energy levels through the crystal-field Hamiltonian derived by Bernal \cite{enrique_optical_1971} for \Er:\Ca through a series of optical spectroscopy measurements:

\begin{equation}
    \mathcal{H}_{\mathrm{cf}} = \sum_{k}^{2,4,6,\dots}\sum_{q}^{-k,...k} B_k^q {O}^q_k \ ,
\end{equation}

\noindent where ${O}^q_k$ are the extended Stevens operators (with $k=2,4,6,\dots$ and $q \in \{-k,\dots, k\}$) \cite{abragam_electron_2012, stevens_matrix_1952}. Because of the $S_4$ symmetry, only certain terms are non-zero, reducing the number of non-zero crystal field parameters $B_k^q$ to six. Their values are taken from \cite{enrique_optical_1971} along with the renormalization factors \cite{erath_crystal_1961}.

Note that the validity of this Hamiltonian is questionable, since the $S_4$ symmetry of the \Er site considered in this work is in fact broken due to the replacement of a nearby calcium by the niobium impurity. On the other hand, the measurements show that the gyromagnetic ratio with $B_0$ applied along the $c$ axis is close to the known value of $\gamma_\parallel$ for $S_4$-symmetric \Er:\Ca, which suggests that the $S_4$-symmetric crystal-field Hamiltonian $\mathcal{H}_{\mathrm{cf}}$ may be a relatively good approximation to the real crystal-field Hamiltonian of this \Er atom.

\paragraph{Zeeman effect}

In presence of a magnetic field, the Kramers doublets are split by the Zeeman effect. The total Hamiltonian describing the \Er $J=15/2$ ground state is thus 

\begin{equation}
    \mathcal{H}_{J} = \mu_B \, g_J\, \mathbf{B_0}\cdot \mathbf{J} + \mathcal{H}_{\mathrm{cf}} , 
\end{equation}

\noindent  where $\mu_B/2\pi = 13.996$~GHz/T is the Bohr magneton and $g_J=6/5$ is the Landé factor of the \Er ground state. 

\paragraph{Effective spin-1/2}

The energy separation between the $Z_1$ and $Z_2$ doublets is approximately 0.6 THz, which corresponds to a temperature of 30~K. Thus, at our working temperature of 10~mK, only the $Z_1$ doublet is populated. This doublet can be described as an effective spin-1/2, with a spin operator $S$, and an anisotropic gyromagnetic tensor,

\begin{equation}
    \mathcal{H}_{S} = \mathbf{B_0}\cdot \bar{\bar{\gamma}}_{\text{Er}}\cdot\mathbf{S},
\end{equation}

\noindent as seen in Fig.\ref{sfig:energy-levels}. The effective gyromagnetic tensor $\bar{\bar{\gamma}}_{\text{Er}}$ is diagonal along the symmetry axes of the crystal, $\gamma_a = \gamma_b = \gamma_\perp$, with $\gamma_\perp /2\pi = - 117.3$~GHz/T, and $\gamma_c = \gamma_\parallel$, with $\gamma_\parallel/2\pi = -  17.45$~GHz/T \cite{antipin_aa_anisotropy_1981}. Defining the $z$ axis as the quantization axis of the spin (which may not coincide with the magnetic field direction, due to the anisotropic gyromagnetic tensor), the Zeeman contribution writes

\begin{equation}
    \mathcal{H}_{S} = \omega_S S_z,
\end{equation}

\noindent with $\omega_S = ||\bar{\bar{\gamma}}_{\text{Er}}\cdot \mathbf{B_0} ||$.

In conclusion, the $Z_1$ Kramers doublet of \Er:\Ca is the effective electron $S=1/2$ paramagnetic center considered throughout the main text. Excited states of the $J=15/2$ manifold contribute to small energy shifts of the $Z_1$ doublet, which are considered in Appendix \ref{app:pseudo_quadrupole_interaction} and give rise to the pseudo-quadrupole effect. Moreover, as discussed earlier, in the presence of a magnetic field, the two $Z_1$ states also possess a small average electric dipole, which is different in each state and which contributes to the Spin-Dependent Quadrupole (see App.\ref{app:electric-dipole_calculation}).

\subsection{Niobium-93}

The \Nb Hamiltonian is 

\begin{equation}
    \mathcal{H}_I =  \omega_I I_z + \mathcal{H}_{Q}
\end{equation}

\noindent In this equation, the first term is the Zeeman energy with $\omega_I = - \gamma_{\text{\Nb}} B_0 $, and $\gamma_{\text{\Nb}} / 2\pi = 10.42$~MHz/T. Here, the quantization axis of the spin, the $z$-axis, coincides with the applied magnetic field $B_0$. The second term is the quadrupolar interaction 

\begin{equation}
    \mathcal{H}_Q =  \mathbf{I} \cdot \bar{\bar{Q}} \cdot \mathbf{I},
\end{equation}

\noindent which describes the electrostatic interaction between the \Nb nucleus quadrupolar moment and the electric field gradients caused by the charges in its immediate vicinity. It is described by a traceless and symmetric tensor $\bar{\bar{Q}}$,

\begin{align}
\begin{split}
    \bar{\bar{Q}} = \begin{pmatrix}
                    Q_{xx} & Q_{xy} & Q_{xz}\\
                    Q_{xy} & Q_{yy} & Q_{yz}\\
                    Q_{xz} & Q_{yz} & Q_{zz}
\end{pmatrix},\\\\
Q_{xx} + Q_{yy} + Q_{zz} = 0.
\label{eq:quadrupole_tensor}
\end{split}
\end{align}

Consequently, the quadrupole tensor is described, in all generality, by $5$ independent quantities. The tensor can be diagonalized in its principal axis basis called $X$, $Y$, $Z$. In this basis, the quadrupolar Hamiltonian writes

\begin{equation}
    \mathcal{H}_Q = Q_X I_X^2 + Q_Y I_Y^2 + Q_Z I_Z^2,
\end{equation}

\noindent with $Q_X + Q_Y + Q_Z = 0$. An alternative form is 

\begin{align}
\begin{split}
        \mathcal{H}_Q = \mathbf{I} \cdot \bar{\bar{Q}} \cdot \mathbf{I} = \frac{C_q}{4I(2I-1)}\cdot \left[3I_{X}^2  - I(I+1)+ \eta\cdot(I_{Y}^2-I_{Z}^2)\right], \\
\end{split}
\end{align}

\noindent  where the quadrupolar interaction strength $C_q$ and biaxiality parameter $\eta$ are introduced. The $5$ quadrupolar parameters are thus the $3$ Euler angles to rotate the $xyz$ into the $XYZ$ basis, plus $(Q_X,Q_Z)$ or $(C_q,\eta)$.

When fitting the nuclear transition frequencies in this work, we consider the measurements with a single orientation of the magnetic field. Under these conditions, the measured NMR frequencies are not sufficient to describe the full quadrupole tensor. It is intuitively clear that the measured spectrum should be invariant under a rotation around the applied field direction $z$, and that therefore only parameters $4$ can be determined at best. This is mathematically evidenced when expressing the quadrupole tensor as a function of rank-2 spherical tensors $T^{(2)}_{m}$, instead of Cartesian coordinates. In the basis of the nuclear spin operators, these tensors take the following form

\begin{equation*}
\begin{split}
T^{(2)}_{0} &= 3 I_z^2 - I(I+1), \\
T^{(2)}_{\pm 1} &= \mp \left( I_z I_{\pm} + I_{\pm} I_z \right), \\
T^{(2)}_{\pm 2} &= I_{\pm}^2 .
\end{split}
\end{equation*}

Through algebraic manipulation of Eq. \ref{eq:quadrupole_tensor} we can rewrite the Hamiltonian as

\begin{equation}
    \mathcal{H}_{Q} = \sum_{m = -2}^2 S_m T^{(2)}_m(\mathbf{I})
    \label{eq:quadrupole_spherical}
\end{equation}

\noindent with

\begin{equation}
\begin{aligned}
S_{0} &= Q_{zz}, \\
S_{\pm 1} &= Q_{zx} \pm i Q_{zy} = |S_1| e^{i\Delta_{\pm1}}, \\
S_{\pm 2} &= \tfrac{1}{2}(Q_{xx} - Q_{yy}) \pm i Q_{xy} = |S_2| e^{i\Delta_{\pm2}}.
\end{aligned}
\end{equation}

A physical rotation around the $z$-axis by an angle $\zeta$ will map the coefficients to

\begin{equation}
S_m \;\mapsto\; e^{-im\zeta}\, S_m,
\end{equation}

Since the energy levels of the Hamiltonian are insensitive to these phase factors and depend only on invariants such as $S_0$, $|S_{\pm 1}|$, and $|S_{\pm 2}|$, the energy spectrum is invariant to rotations along the quantization axis. 

The matrix form of the quadrupole tensor in terms of these invariants is

\begin{align}
\begin{split}
    \bar{\bar{Q}}(\zeta) =&\begin{pmatrix}
-|S_2| \cos(2\zeta+2\Delta)-\frac{S_0}{2}&  |S_2|\sin(2\zeta+2\Delta) & |S_1|\cos(\zeta) \\ 
|S_2| \sin(2\zeta+2\Delta)  & |S_2|\cos(2\zeta+2\Delta)-\frac{S_0}{2} & |S_1|\sin(\zeta) \\
|S_1|\cos(\zeta) & |S_1|\sin(\zeta) & S_0\end{pmatrix} \\\\
\end{split}
\end{align}

where $\Delta$ is the phase difference between $S_{+2}$ and $S_{+1}$

\begin{equation}
\Delta = \Delta_2 - \Delta_1 = \arctan\left(\frac{2Q_{xy}}{Q_{xx}-Q_{yy}}\right) -  \arctan\left(\frac{Q_{yz}}{Q_{xz}}\right)
\end{equation}

For single axis measurements, this leaves $S_0$, $|S_1|$, $|S_2|$, and $\Delta$ to be determined by the fit, whereas $\zeta$ is set to an arbitrary value. 

\subsection{\Nb--\Er hyperfine coupling}

We now consider the \Nb--\Er hyperfine coupling.  In lanthanide ions, the magnetic moment is localized on the atom, and contact hyperfine to the ligands is generally considered negligible. Therefore, we assume that the hyperfine interaction between the \Er and the \Nb is purely magnetic dipolar. Moreover, we assume that this interaction is well described by the point-dipole approximation. Consequently, it is possible to compute the hyperfine tensor coefficients for each lattice site knowing the crystalline parameters. The hyperfine coupling is described by the dipolar Hamiltonian \cite{abragam_electron_2012}

\begin{equation}
    \mathcal{H}_\text{hf} = \frac{\mu_0}{4\pi r^3} 
    \left[ \boldsymbol{\mu}_1 \cdot \boldsymbol{\mu}_2 
    - 3r^{-2} (\boldsymbol{\mu}_1 \cdot \mathbf{r})(\boldsymbol{\mu}_2 \cdot \mathbf{r}), \right]
\end{equation}

\noindent where $\mathbf{r}$ is the vector separating the two spins, with magnetic moment $\boldsymbol{\mu}_1 = g_J \mu_B\ \mathbf{J}$ and $\boldsymbol{\mu}_2 = \gamma_\text{Nb}\ \mathbf{I}$. Expanding the expression in terms of the electron and nuclear spin operators yields

\begin{equation}
\begin{aligned}
\mathcal{H}_\text{hf, J} &= 
g_L \mu_B \ \gamma_\text{Nb} \ \frac{\mu_0}{4\pi r^3} \Big[ 
 J_{x} I_{x} \big( 1 - 3 \sin^2 \theta_r \cos^2 \phi_r \big) \\
&\quad + J_{y} I_{y} \big( 1 - 3 \sin^2 \theta_r \sin^2 \phi_r \big) \\
&\quad + J_{z} I_{z} \big( 1 - 3 \cos^2 \theta_r \big) \\
&\quad - 3 \sin^2 \theta_r \cos \phi_r \sin \phi_r 
\big( J_{x} I_{y} + J_{y} I_{x} \big) \\
&\quad - 3 \sin \theta_r \cos \phi_r \cos \theta_r 
\big( J_{x} I_{z} + J_{z} I_{x} \big) \\
&\quad - 3 \sin \theta_r \sin \phi_r \cos \theta_r 
\big( J_{y} I_{z} + J_{z} I_{y} \big) 
\Big]
\end{aligned}
\end{equation}


\noindent where $\theta_r$ and $\phi_r$ are, respectively, the polar and azimuthal angles of the vector $\mathbf{r}$. This interaction can be written in a more compact form:

\begin{equation}
    \mathcal{H}_{\text{hf,J}} = \mathbf{J}\cdot\bar{\bar{A}}_J\cdot\mathbf{I} \ ,
\end{equation}

\noindent with $\bar{\bar{A}}_J$ the hyperfine tensor in the $J$ representation. This interaction is anisotropic and depends on the relative position and orientation between the two spins. 

In most of this work, we restrict ourselves to the effective spin-1/2 subspace of the $Z_1$ doublet. In this subspace, the hyperfine Hamiltonian is given by 

\begin{equation}
\begin{aligned}
\mathcal{H}_\text{hf} &= 
\frac{\mu_0 \gamma_\text{Nb}}{4\pi r^3} \Big[ 
\gamma_{\perp} S_{a} I_{a} \big( 1 - 3 \sin^2 \theta_r \cos^2 \phi_r \big) \\
&\quad + \gamma_{\perp} S_{b} I_{b} \big( 1 - 3 \sin^2 \theta_r \sin^2 \phi_r \big) \\
&\quad + \gamma_{\parallel} S_{c} I_{c} \big( 1 - 3 \cos^2 \theta_r \big) \\
&\quad - 3 \sin^2 \theta_r \cos \phi_r \sin \phi_r 
\big( \gamma_{\perp} S_{a} I_{b} + \gamma_{\perp} S_{b} I_{a} \big) \\
&\quad - 3 \sin \theta_r \cos \phi_r \cos \theta_r 
\big( \gamma_{\perp} S_{a} I_{c} + \gamma_{\parallel} S_{c} I_{a} \big) \\
&\quad - 3 \sin \theta_r \sin \phi_r \cos \theta_r 
\big( \gamma_{\perp} S_{b} I_{c} + \gamma_{\parallel} S_{c} I_{b} \big) 
\Big] = \mathbf{S}\cdot\bar{\bar{A}} \cdot\mathbf{I}.
\end{aligned}
\label{eq:full-dipole-dipole-hamiltonian}
\end{equation}

\noindent We note that the spin operators are expressed in the crystal axis basis. 

In addition, the hyperfine Hamiltonian can be written in the secular approximation, which consists in keeping only terms proportional to $S_z$. (As already mentioned, the $z$ axis here should be understood as the electron spin quantization axis, which may differ from the applied magnetic field direction). This approximation is justified by the large frequency mismatch between the electron and nuclear spins, which prevents direct energy exchange between the two systems. The secular hyperfine Hamiltonian writes

\begin{equation}
    \mathcal{H}_{\text{hf, sec}} = A_\parallel S_z I_z + A_\perp S_z I_x.
\label{eq:secular_hyperfine}
\end{equation}

Where $x$ is defined as the direction along the perpendicular to the hyperfine interaction. However, the secular hyperfine Hamiltonian in eq. \ref{eq:secular_hyperfine} is written in the quantization basis, which depends on the orientation of $B_0$. For the nuclear spin, the $I_z$ operator will be parallel to the magnetic field but the quantization axis of the \Er spin $S_z$ is given by the anisotropy of its gyromagnetic tensor. In order numerically estimate $A_\parallel$ and $A_\perp$, we project the Hamiltonian in eq. \ref{eq:full-dipole-dipole-hamiltonian} to the spin operators in the quantization basis using the Hilbert-Schmidt inner product

\begin{align}
    A_\parallel = \frac{\text{Tr}\{ H_\text{hf} \times (S_zI_z)^\dagger\}}{\text{Tr}\{ S_zI_z\times(S_zI_z)^\dagger \}},  \quad A_\perp = \frac{\text{Tr}\{ H_\text{hf} \times (S_xI_z)^\dagger\}}{\text{Tr}\{ S_xI_z\times(S_xI_z)^\dagger \}}.
\end{align} 

This formalism can also be used to calculate the other matrix elements of the hyperfine interaction in an arbitrary basis.

\subsection{Lamb shift}

The coupling of a two-level system to a resonator in the weak coupling regime gives rise to enhanced radiative rate (the Purcell effect), but also to frequency shifts (the so-called Lamb shifts). These shifts are induced by the zero-point fluctuations of the microwave field in the resonator, the analogous to the frequency shifts caused by detuned microwave drives, and have been observed in atoms and superconducting qubits~\cite{brune_lamb_1994}. Given the spectral resolution of our measurements, it is necessary to take them into account in our analysis. 

Consider first a spin-1/2 coupled to a resonator in its (vacuum) ground state, with coupling strength $g_0$ and detuning $\Delta$. We consider the so-called bad-cavity limit in which the cavity damping rate is much larger than the coupling, $\kappa \gg g_0$. The spin ground-state $\ket{\downarrow}$, is unperturbed by the Jaynes-Cummings Hamiltonian, since the cavity contains no photon. However, the spin excited state $\ket{\uparrow}$, has its radiative properties modified by the resonator coupling: it acquires an enhanced radiative damping rate (the Purcell effect \cite{purcell_spontaneous_1946}) 

\begin{equation}
\Gamma_\mathrm{R} = \kappa \frac{g_0^2}{\Delta^2 + \kappa^2/4}
\end{equation}

\noindent and it undergoes a frequency shift (the Lamb shift~\cite{julsgaard_measurement-induced_2012}) 

\begin{equation}
\Delta\omega = \Delta \frac{g_0^2}{\Delta^2 + \kappa^2/4}.
\end{equation}

\noindent We now discuss the impact of the Lamb shifts in the case of the coupled \Nb--\Er system. We first note that the \Er ground-state levels $\ket{\downarrow,n}$ do not undergo a frequency shift. The \Er excited state levels $\ket{\uparrow,n}$ on the other hand are shifted by $\Delta \omega_n = g_0^2 \frac{\Delta_n}{\Delta_n^2 + \kappa^2/4}$, where $\Delta_n = \omega_0 - \omega_n$ is the detuning of the EPR-allowed transition from the resonator frequency. The order of magnitude is $\Delta \omega \sim g_0^2 / \kappa \sim 2\pi \times 100\,\mathrm{rad}\cdot \mathrm{s}^{-1}$. The dependence on $n$ causes shifts $\Delta_{n+1} - \Delta_n$ of the NMR transition frequencies $\omega_{n,n+1}^\uparrow $, which therefore must be taken into account in our analysis. 

This is achieved by adding a term

\begin{equation}
    \mathcal{H}_L = \Sigma_{n=0}^{9}  \Delta \omega_n \ket{\uparrow,n}\bra{\uparrow,n}
\end{equation}

\subsection{Complete Hamiltonian}

Combining all the terms, the full Hamiltonian of an \Er:\Ca spin coupled to a \Nb nuclear spin (and to its detection resonator) is

\begin{align}
        \mathcal{H} =\ \mu_B \ g_J\mathbf{B_0}\cdot \mathbf{J} + \mathcal{H}_{\mathrm{cf}} 
        + \omega_I I_z + \mathbf{I}\cdot \bar{\bar{Q}}\cdot \mathbf{I}  
    + \mathbf{J}\cdot\bar{\bar{A}}_J\cdot\mathbf{I} +\mathcal{H}_{\mathrm{L}}
\label{eq:spin_complete_ham}
\end{align}

in the $J=15/2$ modeling, or

\begin{align}
\begin{split}
        \mathcal{H} =  \omega_S S_z + 
        \omega_I I_z + \mathbf{I}\cdot \bar{\bar{Q}}\cdot \mathbf{I}  +
        \mathbf{S}\cdot \bar{\bar{A}} \cdot\mathbf{I} + \mathcal{H}_{\mathrm{L}}
\end{split}
\label{eq:spin_12_hamiltonian}
\end{align}

in the effective-spin-1/2 description. The hyperfine term can be further simplified using the secular approximation~\ref{eq:secular_hyperfine}. 

\section{\Er spin characterization}
\label{app:electron-spin-characterization}


\begin{figure}[t]
    \includegraphics[width=0.6\columnwidth]{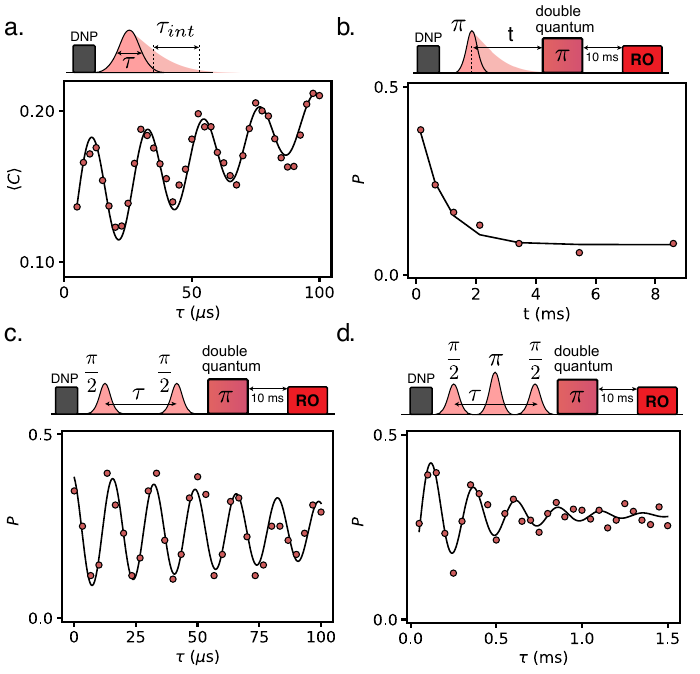}
    \caption{\label{sfig:electron_characterization}
    \textbf{Electron spin characterization}
    \textbf{a.} Electron Rabi oscillation. The dots are measured average counts $\langle C \rangle$ over $\tau_{int}$ plot as a function of the excitation pulse duration, fitted to an exponentially decaying cosine with linearly increasing background (solid line).
    \textbf{b.} Electron fluorescence. The probability to find the electron spin in the excited state is shown as a function of the delay after the excitation pulse. The solid line is an decaying exponential fit which yields a relaxation time $T_1=0.5$~ms
    \textbf{c \& d.} Electron Ramsey \& Hahn echo measurements. The probability to find the electron spin in the excited state is shown as a function of the interpulse delay $\tau$. The solid line is an exponentially decaying cosine fit (line), yielding $T_2^*=0.10$~ms and $T_2=0.35$~ms.
    }
    \label{fig:electron_characterization}
\end{figure}


We characterize the \Er's properties through pulsed EPR measurements. 
Rabi oscillations of the  \Er are shown in Fig.\ref{sfig:electron_characterization}b. The approximately linear increase of $\langle C\rangle$ with pulse duration is attributed to excitation of other emitters coupled to the resonator, either charged TLS or spins, as already observed in related work \cite{wang_single-electron_2023, albertinale_detecting_2021}. 

The \Er spin can also be characterized by using the \Nb nuclear spin as a memory, as demonstrated for NV centers in diamond with proximal $^{13}\mathrm{C}$ nuclear spins~\cite{jiang_repetitive_2009}. To achieve this, the \Er pulse sequence is followed by a $\pi$-pulse on the double-quantum transition (see App.~\ref{app:zero-double-quantum-transitions} and Fig.\ref{sfig:double-quantum}), which converts the $\ket{\uparrow,0}$ population into population of the $\ket{\downarrow,1}$, followed by nuclear spin readout. The method yields better signal-to-noise ratio than simply measuring the fluorescence, since the nuclear spin state readout single-shot. Single-shot \Er readout is used to measure the \Er relaxation time, $T_1=0.5$~ms, the Free-Induction-Decay time $T_2^*=0.10$~ms, and the Hahn echo decay time $T_2=0.35$~ms (Fig.\ref{sfig:electron_characterization}c, d \& e respectively).

Since the relaxation rate is mostly dominated by the Purcell rate, we can estimate the coupling coupling constant $g_0$ between the \Er spin and the resonator, ~\cite{wang_single-electron_2023}

\begin{equation}
     \frac{g_0}{2\pi}= \frac{1}{2\pi} \sqrt{\frac{\kappa}{4T_1}} = 9.3 \text{~kHz}.
\end{equation}

\noindent From this estimate, we infer that the \Er lies $\approx 50\,$nm below the sample surface, as seen from the computed $g_0$ spatial dependence (as calculated in previous work \cite{travesedo_all-microwave_2025}). Note also that the measurements of Fig.\ref{fig3}b may indicate that non-radiative relaxation is not negligible for this ion, so that $g_0$ might be slightly over-estimated.


\section{Zero- and double-quantum transitions}
\label{app:zero-double-quantum-transitions}

In this section, we derive analytical equations for the matrix elements of the \Nb zero- and double-quantum transitions. We emphasize that state $\ket{0}$ is close to $\ket{m_s=+9/2} $, $\ket{1}$ to $\ket{m_s=+7/2}$, etc. Therefore double-quantum transitions are $\ket{\downarrow, n+1} \leftrightarrow \ket{\uparrow, n}$, and zero-quantum transitions are transitions are $\ket{\downarrow, n} \leftrightarrow \ket{\uparrow, n+1}$. We find good agreement with numerical diagonalisation of the Hamiltonian. We then measure the double-quantum transition spectrum, and Rabi oscillation. We finally use these results to obtain $A_\perp$.

\subsection{Hamiltonian}

We first derive an approximate Hamiltonian in which the quadrupole interaction takes a simpler form. For that, we rely on the finding (see main text) that one of the principal axis of the quadrupole ($Z$) lies along the crystalline $c$-axis. The other two axis lie in the $(a,b)$-plane and the $X$ axis makes an angle $\alpha_Q=-10^\circ$ with respect to the crystalline $a$-axis.





In the principal axis basis, the quadrupole Hamiltonian can be expressed as

\begin{align}
\begin{split}
        \mathcal{H}_Q = \mathbf{I} \cdot \bar{\bar{Q}} \cdot \mathbf{I} = \frac{C_q}{4I(2I-1)}\cdot \left[3I_{X}^2 - I(I+1) + \eta\cdot(I_{Y}^2-I_{Z}^2)\right] \\
\end{split}
\end{align}

\noindent We now wish to express this Hamiltonian in the $x, y, z$ coordinate system. In this frame, $z$ corresponds to the direction of $B_0$, which is approximately parallel to the $c$-axis, and $x$ to the direction of the perpendicular hyperfine coupling, which makes an angle of $\alpha_\text{hf}=45^\circ$ with the crystalline $a$-axis. In this frame, the quadrupole Hamiltonian takes the following form

\begin{align}
\begin{split}
        \frac{\omega_Q'}{6} &\left[3I_{z}^2 + \frac{\eta'}{2}(e^{-2i\alpha}I_+^2-e^{2i\alpha}I_{-}^2)\right] + \text{cnt}. \\
        \omega_Q' &= -\frac{\eta+1}{2} \cdot \frac{3C_q}{2I(2I-1)} = 2\pi \times -712~\text{kHz}\\
        \eta' &= \frac{\eta-3}{1+\eta} = 1.25 \\
        \alpha &= \alpha_\text{hf} - \alpha_Q.
\end{split}
\end{align} 
 
\noindent Therefore, the Hamiltonian of the system can be approximated by 

\begin{align}
\begin{split}
        \label{seq:approx_hamiltonian}
        \mathcal{H} =\ \omega_S\cdot S_z + \omega_I\cdot I_z + A_\perp S_z I_x + A_\parallel S_z I_z\ + \frac{\omega_Q'}{6} \left[3I_{z}^2 + \frac{\eta'}{2}(e^{-2i\alpha}I_+^2-e^{2i\alpha}I_{-}^2)\right]\,
\end{split}
\end{align}

\noindent where the nuclear spin operators are all written the basis defined by the direction of the perpendicular hyperfine coupling.


\subsection{Matrix elements of the nuclear-spin-flipping transitions}

We consider as a perturbation all terms that do not commute with $S_z$ and $I_z$ in equation \ref{seq:approx_hamiltonian}, that is, the perpendicular hyperfine term $A_\perp S_z I_x$ and the non-diagonal quadrupole terms. As such we can write the Hamiltonian as 

\begin{align}
\begin{split}
        \mathcal{H} &= \mathcal{H}_0+V \\
        \mathcal{H}_0 &= \omega_S\cdot S_z + \omega_I\cdot I_z + A_\parallel S_z I_z +  \omega_Q'/2\cdot I_z^2 \\
         V & = A_\perp S_z I_x + \frac{\omega_Q'\eta'}{12}\cdot(e^{-2i\alpha}I_+^2-e^{2i\alpha}I_{-}^2),
\end{split}
\end{align}

\begin{figure}[t]
    \includegraphics[width=\columnwidth]{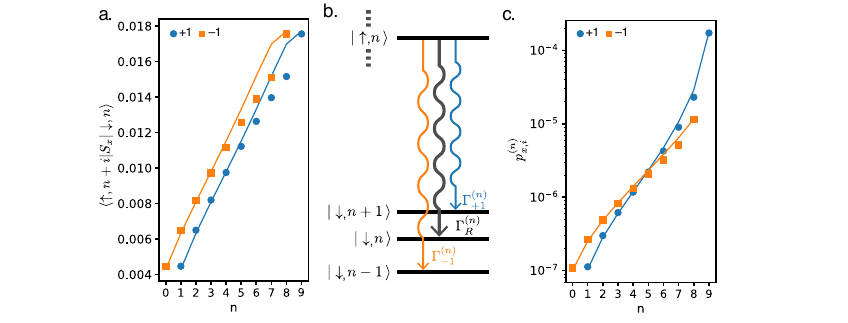}
    \caption{\label{sfig:mixing}
    \textbf{Mixing terms and cross-relaxation probability.} Throughout this figure, dots are obtained through the analytical expressions derived in the text and lines via numerical simulation using experimental parameters.
    \textbf{a} Matrix elements $\langle \uparrow, n + i | S_x | \downarrow, n \rangle$ as a function of $n$ for $i=+1,-1$ (blue and orange respectively). 
    \textbf{b}. Energy level diagram (black horizontal lines) showcasing all possible relaxation paths (colored lines) and their respective rates $\Gamma^{(n)}_i$.
    \textbf{c} Cross-relaxation probability $p^{(n)}_{x,i}$ as a function of $n$ for $i=+1,-1$ (blue and orange respectively).
    }
\end{figure}

\noindent where $V$ acts as a perturbation to the diagonal Hamiltonian $\mathcal{H}_0$. The eigenstates of the unperturbed system are $\ket{m_s m_I}$, where $m_s\in\{\uparrow\,\equiv1/2 ,\; \downarrow\,\equiv-1/2\}$ and $m_I \in \{-9/2,\dots,9/2\}$ are the electron and nuclear spin quantum numbers. The hyperfine perturbation and quadrupole perturbations will mix terms with $m_I \pm 1$ and $m_I \pm 2$ respectively. At first order, the perturbed states will be

\begin{equation}
\label{seq:first-order-wfn}
    \ket{\psi_{m_S,m_I}} \approx \ket{m_S, m_I} + \sum_{i\in\{\pm1\pm2\}} M_i^{(m_S, m_I)} \cdot \ket{m_S, m_I + i}
\end{equation}

\noindent where $M_i^{(m_S, m_I)}$ correspond to the mixing terms between the energy levels and are given by

\begin{align}
\begin{split}
\label{seq:mixing_coefficients}
    &M_{\pm1}^{(m_S,m_I)} =  \mp\frac{C^\pm_{m_I} \cdot A_\perp m_S / 2}{\omega_I+A_\parallel m_S + \omega'_Q (m_I\pm1/2)}, \\
    &M_{\pm2}^{(m_S,m_I)} = \mp\frac{e^{\mp2i\alpha}D^\pm_{m_I} \cdot \omega_Q'\eta'/12}{2(\omega_I+A_\parallel m_S + \omega'_Q (m_I\pm1))}.\\ \\
    &C^\pm_{m_I} = \bra{m_I+1}I_\pm\ket{m_I} = \sqrt{I(I+1)-m_I(m_I\pm1)},\\
    &D^\pm_{m_I} = \bra{m_I+2}I_\pm^2\ket{m_I} = C^\pm_{m_I} \, C^\pm_{m_I+1}\,.
\end{split}
\end{align}

\noindent Given that the value of the mixing terms is not negligible, the nuclear spin magnetic numbers are not good quantum numbers anymore, and we introduce the notation $\ket{m_S, n} = \ket{\psi_{m_s, 9/2-m_I}}$ with $n\in\{0,\dots,9\}$ in ascending order of energy. 

The mixing terms create differences between the nuclear wavefunctions depending on the state of the electron spin, evidenced by the appearance of the $m_S$ term in the $M_{i}$ expressions. It is precisely in this difference that the origin of the zero- and double-quantum transitions lies. The oscillator strength of these transition is given by the matrix element of the $S_x$ operator between levels with $\Delta n =\pm1$. To first order in perturbation theory, these matrix elements are 

\begin{align}
\begin{split}
    &\bra{\downarrow, n}S_x\ket{\uparrow, n} \approx 1/2\\
    &\bra{\downarrow, n\pm1}S_x\ket{\uparrow, n} \approx \mp\frac{1}{4}
\frac{A_\perp C^\pm_{n}}{\omega_I+\omega'_Q\!\left(n\pm\tfrac12\right)}
, \\[6pt]
\end{split}
\end{align}


The elements are shown in Fig.\ref{sfig:mixing}.b, both the analytical expressions and the direct diagonalisation (dots and lines resp.). 


\subsection{Cross-relaxation rates}

Using the calculated matrix elements we now proceed to give an estimation of the relaxation rate of the electron spin through all the available pathways. As noted before, the main relaxation channel for the electron spin leaves the nuclear spin invariant. Moreover, the oscillator strength for these transitions is approximately 1/2 no matter the state of the nuclear spin. However, due to the hyperfine interaction the frequency of these transitions ($\omega_0, \dots,\omega_9$)  is nuclear spin dependent. As such, when the resonator is centered at $\omega_0$ the Purcell factor for each of them will be different. The electron relaxation rate when the nuclear spin is in state $n$ is thus given by
\begin{equation}
    \Gamma_n = \frac{\kappa g_0^2}{\kappa^2/4 + (\omega_\mathrm{r} - \omega_n)^2} =  \Gamma_0 \cdot \frac{1}{1 + 4\frac{(\omega_\mathrm{r} - \omega_n)^2}{\kappa^2}}
\end{equation}
where $\Gamma_0 = 4g_0^2/\kappa = 1.72\cdot10^{-3}$~s$^{-1}$ is the electron decay rate when $n=0$, which is experimentally obtained (see App.\ref{app:electron-spin-characterization}). For a cross-relaxation where the state of the nuclear spin is changed from $n$ to $n+i$ and with transition frequency $\omega_{n}^{(i)}$, the decay rate is obtained by rescaling the relaxation rate given by the Purcell effect by the oscillator strength

\begin{equation}
    \Gamma^{(i)}_{n} = \frac{\kappa g_0^2}{\kappa^2/4 + (\omega_\mathrm{r} - \omega_n^{(i)})^2} \cdot \frac{|\bra{\downarrow, n+i}S_x\ket{\uparrow, n}|^2}{|\bra{\downarrow, n}S_x\ket{\uparrow, n}|^2}  = 2\,\Gamma_0 \cdot \frac{|\bra{\downarrow, n+i}S_x\ket{\uparrow, n}|^2}{1 + 4\frac{(\omega_\mathrm{r} - \omega_{n}^{(i)})^2}{\kappa^2}}
\end{equation}

The total relaxation rate for $\ket{\uparrow n}$ is given by $\Gamma^{(n)} = \Gamma^{(n)}_R +\sum_i\Gamma^{(n)}_i$. Figure \ref{sfig:mixing}.c illustrates all the possible relaxation paths through an energy diagram. We naturally introduce the cross-relaxation probability

\begin{equation}
    p^{(n)}_{x,i} = \frac{\Gamma^{(n)}_i}{\Gamma^{(n)}} \approx \frac{\Gamma^{(n)}_i}{\Gamma^{(n)}_R} = \frac{2|\bra{\downarrow n+i}S_x\ket{\uparrow n}|^2}{1 + 4\frac{(\omega_\mathrm{r} - \omega_{n}^{(i)})^2}{\kappa^2}}.
\end{equation}

which are shown as a function of $n$ in figure \ref{sfig:mixing}.d for $i\in\{\pm1\}$. These vastly different cross-relaxation rates explain why low-$n$ states appear more stable in the time trace of Fig.1c in the main text. To maximize the fidelity of the readout scheme, all nuclear spin state measurements are performed in the $n=0$ state, in which the cross-relaxation probability is minimized. 

\subsection{Spectroscopy of the Double-quantum transitions and estimate of $A_\perp$}

\begin{figure}[t]
    \includegraphics[width=\textwidth]{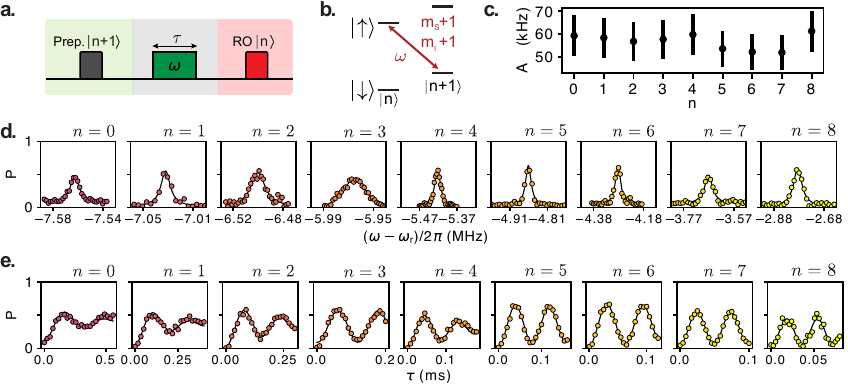}
    \caption{\label{sfig:double-quantum}
    \textbf{Double-quantum transitions.}  \
    \textbf{a.} Pulse sequence used to measure double-quantum transitions. 
    \textbf{b.} Level diagram of double-quantum transitions. 
    \textbf{c.} Calculated $A_\perp$ using the Rabi frequency measured for all double-quantum transitions.
    \textbf{d.} Double-quantum transition spectroscopy for $n\in\{0,...,8\}$. The solid lines are lorentzian fits.
    \textbf{e.} Double-quantum transition Rabi oscillations. The solid lines are fits using a cosine with an decaying exponential envelope.
    }
\end{figure}

We now proceed to measure all double-quantum transitions (see Figure \ref{sfig:double-quantum}). We first prepare the state $\ket{\downarrow,n+1}$, then apply a monotone microwave pulse of frequency $\omega$ and duration $\tau$. We wait 5 ms for the \Er spin to relax to the ground state before reading out $\ket{n}$. The Rabi frequency is given by



\begin{equation}
\begin{split}
    \Omega_n^{(\text{DQ})} &= \mathcal{A} \cdot \langle \uparrow, n | S_x | \downarrow, n+1 \rangle \\
    \mathcal{A} &=|\bar{\bar{\gamma}}_\text{\Er} \cdot \mathbf{B}_{1}| \cdot \frac{1}{\sqrt{1 + 4\frac{(\omega_\mathrm{r}-\omega_{n}^{(-)})^2}{\kappa^2}}}
\end{split}
\end{equation}

\noindent where the amplitude of the drive is filtered by the resonator.

To quantify the amplitude of the microwave drive, we relate it to the electron Rabi frequency $\Omega_e$ that we measure at a reference pulse amplitude. For a resonant electron spin transition,

\begin{equation}
    \Omega_{e} = |\bar{\bar{\gamma}} \cdot \mathbf{B}_{1}| \cdot \bra{\uparrow, n} S_x \ket{\downarrow, n} = \tfrac12 |\bar{\bar{\gamma}} \cdot \mathbf{B}_{1}|, 
\end{equation}

Using $\Omega_e$ as a calibration point, we express the drive amplitude for any pulse as a relative factor $\alpha$. The corresponding sideband Rabi frequency is then 

\begin{equation}
\begin{split}
    \Omega_n^{(\text{DQ})} =\; \frac{\Omega_{e}}{\alpha} \cdot \frac{2{\langle \uparrow, n | S_x | \downarrow, n+1 \rangle}}{\sqrt{1 + 4\frac{(\omega_\mathrm{r}-\omega_{n}^{(-)})^2}{\kappa^2}}}
    =\; \frac{A_\perp C^-_{n}}{2\alpha\cdot\left(\omega_I+\omega'_Q\!\left(n+\tfrac12\right)\right)}\frac{\Omega_{e}}{\sqrt{1 + 4\frac{(\omega_\mathrm{r}-\omega_{n}^{(-)})^2}{\kappa^2}} } 
\end{split}
\end{equation}

We use this expression to estimate the value of the perpendicular hyperfine coupling independently for each transition which averages to $A_\perp=55(8)$~kHz. We note that the error is mainly dominated by the uncertainty of the resonator losses $\kappa$.


We note that the zero- and double-quantum Rabi rates are of the same order of magnitude as the electron-spin $T_2^*$ at the maximum microwave amplitude available in our setup. Under these conditions, high-fidelity $\pi$-pulses cannot be reliably defined, as they are dominated by coherent errors. This limitation is especially significant for the low-$n$ transitions, whose matrix elements are substantially smaller than those of their high-$n$ counterparts.

\section{\Nb site assignment}
\label{app:angular_dependence}

As described in Appendix \ref{app:spin_hamiltonian}, the \Er ion and the \Nb nuclear spin are coupled by the magnetic dipolar interaction, which depends on the relative position of the two dipoles. In this section, we use the measured value of the hyperfine couplings to determine the \Nb position relative to the \Er ion. 

\begin{figure}[t]
    \includegraphics[width=\textwidth]{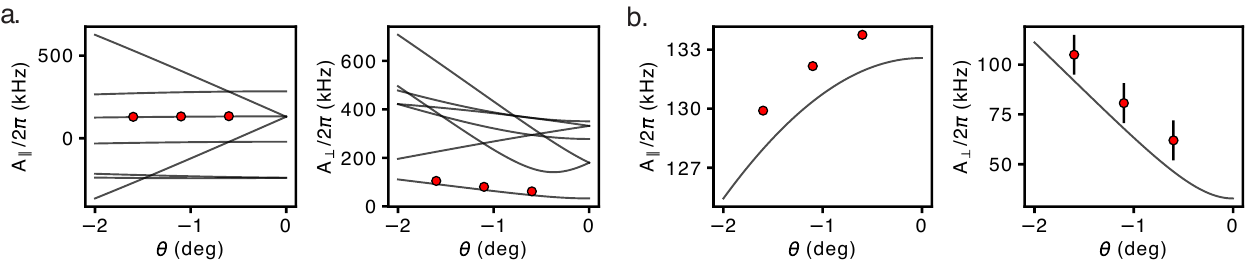}
    \caption{\label{sfig:hyperfine_angle_dependence}
    \textbf{\Nb hyperfine coupling as a function of $\theta$}.
    \textbf{a.} Calculated hyperfine couplings $A_{\parallel}$ and $A_{\perp}$ (left and right panels, respectively) as a function of the in-plane angle $\theta$ between \Er and \Nb for all ten neighboring W sites in the first unit cell (black lines), together with the experimentally measured values (red circles). The numerical estimates take into consideration the out-of-plane angle $\beta_0$.
    \textbf{b.} Magnified view of the relevant region only showing the coupling for a type 3 position. Error-bars for the $A_\parallel$ measurement are hidden by the marker.
    }
\end{figure}

Figure \ref{sfig:hyperfine_angle_dependence}.a plots the measured values of $A_\parallel$ and $A_\perp$ along with with the numerical estimates for all 10 possible W positions in the first unit cell. Due to symmetry, the coupling strength overlaps for the type 3 sites and pairwise for the type 1 site. However, the out-of-plane angle $\beta_0$ breaks the degeneracy between the four type 2 spins. The measured data quantitatively agrees with the coupling for a type 3 position, which lies directly along the $c$-axis. Due to the degeneracy of type 3 sites, we cannot distinguish between the two possible configurations. Figure \ref{sfig:hyperfine_angle_dependence}.b shows a magnified version of the plot. We note a systematic shift for the values of $A_\parallel$ of less than 1\% with respect to the measured values, which may be due a breakdown of the point-dipole approximation, or to a slight shift of the \Nb position with respect to the \W site, possibly also impacted by strain caused by differential thermal contraction of the \Nb thin-film and the \Ca substrate \cite{pla_strain-induced_2018,billaud_electron_2025}. 

\section{Dynamic nuclear polarization of the \Nb and the \W bath}
\label{app:dnp}

\begin{figure}[t]
    \includegraphics[width=\textwidth]{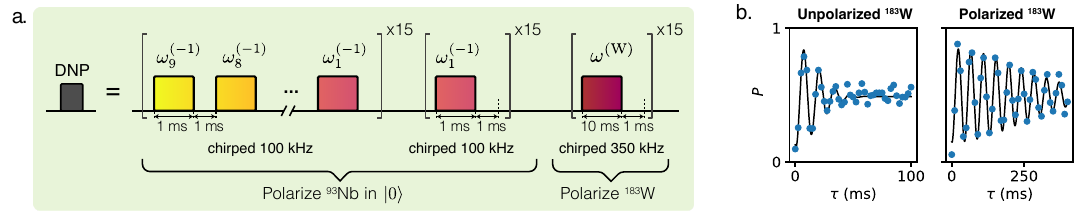}
    \caption{\label{sfig:sequence}
    \textbf{DNP pulse sequence and coherence of \Nb with and without \W polarization}.
    \textbf{a.} Pulse sequence for the DNP. The sequence is separated in two distinct parts: the preparation of the \Nb spin in $\ket{0}$ and the polarization of the \W nuclear spin bath via solid effect. Pulse durations are noted as black arrows and the chirp is explicitly stated under every pulse.
    \textbf{b.} Ramsey sequence with and without \W polarization pulses. Measurement data (dots) are fit to an exponentially decaying (line) which yields $T_2^*=22$~ms and $T_2^*=331$~ms for the unpolarized and the polarized case respectively.
    }
\end{figure}

In this section we describe the sequences used to deterministically prepare the state of the \Nb spin to $\ket{0}$ before every pulse sequence using Dynamic Nuclear Polarization (DNP). We also detail how we use DNP to reduce the spectral noise of neighboring \W nuclear spins on the \Nb coherence.

Deterministic state preparation is a key capability in the control of any quantum system. In previous work~\cite{travesedo_all-microwave_2025}, we demonstrated that pumping the zero- and double-quantum transitions can be used to polarize strongly coupled individual \W\ nuclear spins, using a technique known as solid-effect DNP~\cite{abragam_a_proctor_w_nouvelle_1958}. Here, we extend this approach to the \Nb\ nuclear spin with $I=9/2$ by sequentially driving all double-quantum transitions (see App.~\ref{app:zero-double-quantum-transitions}), starting from $n=8$ down to $n=0$, using 1~ms microwave pulses. Each pulse is followed by a 1~\textmu s delay to allow the electron spin to relax to its ground state. Because of their small matrix elements and large detuning relative to the allowed transitions, the zero- and double-quantum transitions experience large AC-Zeeman shifts~\cite{travesedo_all-microwave_2025}, causing their resonance frequencies to be sensitive to the applied microwave amplitude. To compensate for microwave-power fluctuations and spectral drift, all pulses are chirped by 100~kHz. Due to the low Rabi rates of the sideband transitions, a single pulse is insufficient to achieve full population transfer. We therefore apply 15 repetitions of the pumping sequence to ensure reliable preparation of the \Nb\ nuclear spin. An additional 15 pulses are applied to the $n=0$ transition, which has the smallest matrix element and was experimentally found to be particularly difficult to polarize.

The nuclear spin bath, consisting of neighboring \W\ nuclear spins, significantly contributes to the spectral drift of the \Nb\ transition frequencies. To mitigate this effect, we polarize the spin bath using solid-effect DNP by applying microwave pulses resonant with the double-quantum transitions of the coupled \W\ spins. To first order, the double-quantum transition is detuned from the electron-spin transition by the Larmor frequency of the target nuclear spin, which for \W\ at $ 446$~mT is $\sim 800$~kHz~\cite{travesedo_all-microwave_2025}. To be able to effectively drive weakly coupled \W nuclear spins, we employ 40~ms pulses separated by 10~ms delays to allow the microwave lines to cool, and repeat this sequence 15 times. The strong pulse amplitudes induce a substantial AC-Zeeman shift and we account for this by chirping the pulse frequency by an amount of 350~kHz around $\omega^{\mathrm{(W)}}/2\pi = 725$~kHz.

The complete DNP sequence used in this work is shown in Fig.\ref{sfig:sequence}a. Note that the \W\ polarization sequence needs to be applied after the \Nb\ preparation, since the electron-spin allowed transition depends on the \Nb\ state, and the \W\ double-quantum transition is correspondingly detuned relative to this transition.

To demonstrate the impact of the \W polarization, Fig.\ref{sfig:sequence}b shows a \Nb Ramsey measurement with and without \W polarization. The \Nb coherence time $T_2^*$ without \W polarization is $26$\,ms, and it increases up to 331~ms upon polarizing the \W bath.


\section{Microwave stimulated Raman Driving}
\label{app:stimulated_raman_driving}

\begin{figure}[t]
    \includegraphics[width=\textwidth]{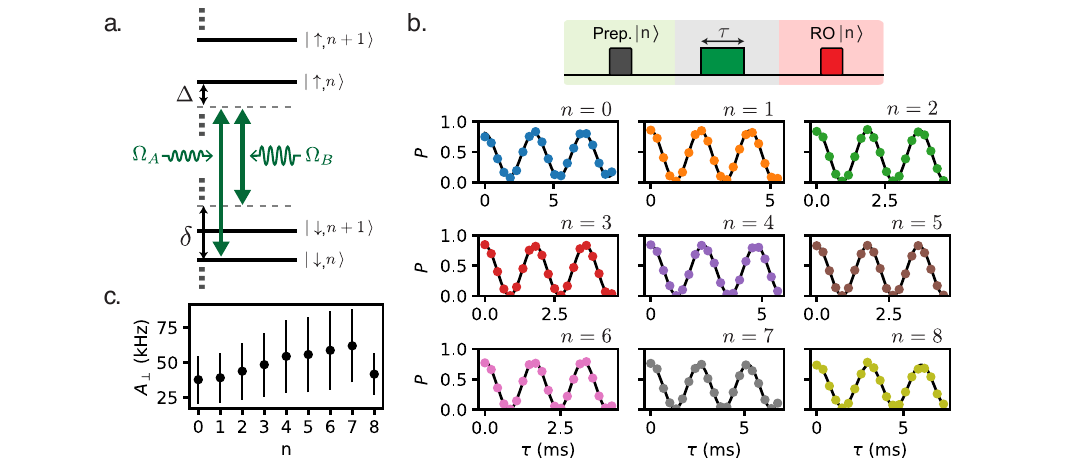}
    \caption{\label{sfig:rabis}
    \textbf{Rabi oscillations for all NMR transitions}.
    \textbf{a.} Reduced energy level scheme with driving pulses. Only the four relevant energy levels are represented. Green arrows show the two Raman driving pulses with amplitudes $\Omega_\text{A}$ and $\Omega_\text{B}$. The detuning between the two drives is $\delta$ and the detuning between the EPR transition $\ket{\downarrow, n} \leftrightarrow \ket{\uparrow, n}$ and the first microwave drive is given by $\Delta$.
    \textbf{b.} Rabi oscillations for all NMR transitions. \textit{(top)} Pulse sequence. We prepare the state $\ket{\downarrow, n}$ followed by a resonant Raman pulse with duration $\tau$. The population of nuclear state $\ket{n}$ subsequently readout.
    \textit{(bottom}. Measured Rabi oscillations (colored dots) and cosine fit (solid dark line) as a function of $\tau$. The amplitude of the drives $\Omega_A$ and $\Omega_B$ was calibrated to give similar Raman Rabi frequencies.
    \textbf{c.} Estimated $A_\perp$ for all NMR transitions as a function of $n$. Solid dots represent the estimate while the dashed line plots the average of all measurements.
    }
\end{figure}

In this section we derive the expressions for the Rabi frequency of a Raman drive and use them to estimate the value of $A_\perp$. To coherently transfer the population between adjacent nuclear spin states, $\ket{\downarrow, n}$ and $\ket{\downarrow, n+1}$, we use microwave stimulated Raman driving. In this scheme, two detuned microwave pulses, with amplitudes $\Omega_A$ and $\Omega_B$, are simultaneously applied. This technique is useful when one is interested in operating between two \textit{apriori} uncoupled levels, provided that both are coupled to a third one. In our particular case, both $\ket{\uparrow, n}$ and $\ket{\uparrow, n+1}$ are significantly coupled to the levels of interest, which results in a 4-level interaction, as depicted in Fig.\ref{sfig:rabis}.a. Under a Raman drive, there are two relevant frequencies: $\delta$, the frequency difference between the two pulses, and $\Delta$, the frequency difference between the first drive and the EPR transition of reference, in the case of the diagram $\omega_n$. 

To quantify the strength of the drive on the \Er we relate the amplitude of the stimulated Raman pulses with the electron Rabi frequency at a specific pulse amplitude $\Omega_e$, as defined in App.\ref{app:zero-double-quantum-transitions}

\begin{equation}
\begin{split}
    \Omega_{A} = \frac{\Omega_{e}}{\alpha_{A}} \cdot \frac{1}{\sqrt{1 + 4\frac{(\omega_\mathrm{r}-(\omega_n-\Delta))^2}{\kappa^2}}}, \\
    \Omega_{B} = \frac{\Omega_{e}}{\alpha_{B}} \cdot \frac{1}{\sqrt{1 + 4\frac{(\omega_\mathrm{r}-(\omega_n-\Delta-\delta))^2}{\kappa^2}}},
\end{split}
\end{equation}

where $\alpha_{A,B}$ is a linear scaling factor to account for differences in pulse amplitudes. The second term of the expression accounts for the resonator filtering of the microwave amplitude. Given that there are two coupled levels ($\ket{\uparrow, n}$ and $\ket{\uparrow, n+1}$) we consider two paths, each with its own Rabi frequency,

\begin{equation}
\begin{split}    
    \Omega_{\mathrm{Ram}} =&\; \Omega_{\mathrm{Ram},1} +\Omega_{\mathrm{Ram},2} \\ \\
    \Omega_{\mathrm{Ram},1} =&\; 
    \frac{\Omega_A\Omega_B}{2\Delta} \cdot \frac{\bra{\uparrow, n+1} S_x \ket{\downarrow, n}}{\bra{\uparrow, n} S_x \ket{\downarrow, n}} 
    \approx -\frac{\Omega_A\Omega_B}{4\Delta} \frac{A_\perp C^+_{n}}{\omega_I+\omega'_Q\!\left(n+\tfrac12\right)} \\ \\
    \Omega_{\mathrm{Ram},2} =&\; 
    \frac{\Omega_A\Omega_B}{2\left(\Delta + \omega_n\right)} \cdot \frac{\bra{\uparrow, n} S_x \ket{\downarrow, n+1}}{\bra{\uparrow, n+1} S_x \ket{\downarrow, n+1}}
    \approx + \frac{\Omega_A\Omega_B}{4\left(\Delta + \omega_n\right)} \frac{A_\perp C^+_{n}}{\omega_I+\omega'_Q\!\left(n+\tfrac12\right)}
\end{split}
\end{equation}

The values of the matrix elements were derived in App.\ref{app:zero-double-quantum-transitions}. We note that the maximum value for the total Rabi frequency $\Omega_{\text{Ram}}$ is given for $\Delta=-\omega^{(\uparrow)}_{n, \ n+1}/2$, as driving in between the states compensates the difference in sign between the two drive amplitudes, resulting in

\begin{equation}
    \Omega_{\mathrm{Ram}} = \frac{\Omega_A\Omega_B}{\omega^{(\uparrow)}_{n, \ n+1}} \frac{A_\perp C^+_{n}}{\omega_I+\omega'_Q\!\left(n+\tfrac12\right)} \approx A_\perp C^+_{n}\frac{\Omega_A\Omega_B}{\omega_{n, \ n+1}^2},
\end{equation}

\noindent where we have used that 

\begin{equation}
\omega_{n, \ n+1} =\omega_I+\omega'_Q\!\left(n+\tfrac12\right) = \frac{\omega^{(\downarrow)}_{n, \ n+1} + \omega^{(\uparrow)}_{n, \ n+1}}{2} \approx \omega^{(\uparrow)}_{n, \ n+1} \approx \omega^{(\downarrow)}_{n, \ n+1}.
\end{equation}

Figure \ref{sfig:rabis}.b shows the Rabi oscillations for all NMR transitions along with the pulse sequence. The data was fit to a cosine and the resulting frequency $\Omega^{\mathrm{fit}}_{\mathrm{Ram}}(n)$ was used to estimate $A_\perp$ since

\begin{equation}
    A_\perp =\frac{\Omega_{\mathrm{Ram}}}{C^+_{9/2-n}}\cdot \frac{\omega_{n, \ n+1}^2}{{\Omega_A\Omega_B}}
\end{equation}

Figure \ref{sfig:rabis}.c plots the estimated $A_\perp$ as a function of $n$. The data points are centered around 50(11)~kHz. The value matches quantitatively with the one measured through the sideband driving (see App.\ref{app:zero-double-quantum-transitions}). 
The remaining positive trend cannot be attributed to second order effects, since a comparison between the analytical and measured matrix elements gives a relative difference on the order of 2\% (see Fig.\ref{sfig:mixing}b). Given that the main origin of the uncertainty of the measurement is the resonator linewidth, we attribute the origin of this trend to small miss-characterization of $\omega_0$ and $\kappa$ at the time of the measurement. The resonator plays a crucial role on the filtering of the microwave amplitudes and its parameters can slowly change over time.

\section{Identification of the $^{93}$Nb atom}
\label{app:nb_id}

We use the gyromagnetic factor of the nuclear spin to identify its chemical nature. As discussed in the main text, this quantity can be first estimated from the $\ket{4} \leftrightarrow \ket{5}$ transition, $\frac{ \omega^{(\downarrow)}_{45} + \omega^{(\uparrow)}_{45}}{2}$, as it is first order insensitive to quadrupole shifts. A more precise estimate is obtained using the Larmor frequency $\omega_I$ from the final fit $\omega_I /2\pi = -4735.52$\,kHz, which in combination with the applied field $B_0 = 0.44627$\,T yields $\gamma/2\pi = 10.61$\,MHz/T. 
By comparison with the tabulated values of the gyromagnetic ratio of stable nuclei with spin-9/2 (see Table \ref{tab:nuclear_spin}) we identify the spin as a \Nb. The remaining difference with the tabulated value is likely due to a combination of a pseudo-nuclear-Zeeman shift caused by the \Er, and/or to chemical shifts.

\begin{table}[h!]
\centering
\begin{tabular}{lccccccc}
\hline
Isotope & $^{73}$Ge & $^{83}$Kr & $^{87}$Sr & $^{93}$Nb & $^{113}$In & $^{115}$In & $^{209}$Bi \\ \hline
$\gamma / 2\pi$ (MHz/T) & $-1.489$ & $-1.644$ & $-1.851$ & $+10.452$ & $+9.365$ & $+9.386$ & $+6.962$ \\ \hline
\end{tabular}
\caption{Gyromagnetic ratios of stable nuclei with spin $I=9/2$.}
\label{tab:nuclear_spin}
\end{table}

\section{Ramsey and Hahn echo measurements}
\label{app:full_ramsey_data}

\begin{figure}[t]
    \includegraphics[width=\textwidth]{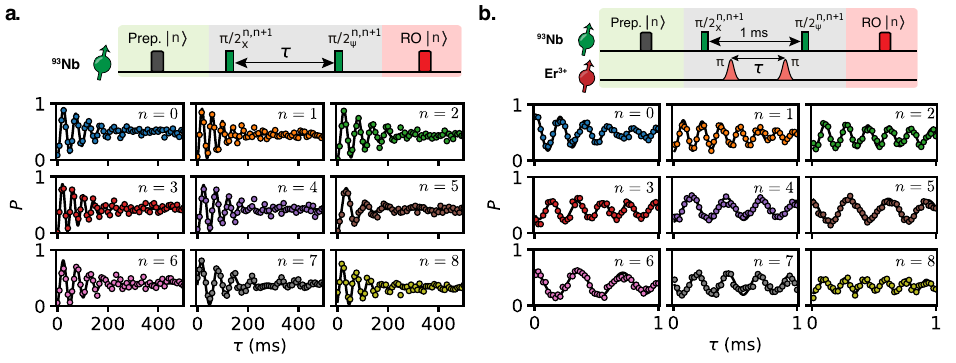}
    \caption{\label{sfig:coherences}
    \textbf{Ground and excited state Ramsey measurements for all NMR transitions}.
    \textbf{a.} Ground state Ramsey measurements. The probability to find the \Nb in state $\ket{n}$ is plotted as a function of interpulse delay $\tau$. The solid lines are cosine fits with a Gaussian decaying envelope. 
    \textbf{b.} Excited state Ramsey measurements. The probability to find the \Nb in state $\ket{n}$ is plotted as a function of interpulse delay $\tau$. The solid lines are cosine fits with an exponentially decaying envelope. 
    }
\end{figure}

 Ground- and excited-state Ramsey measurement for all NMR transitions were performed in an interleaved manner (see Fig.\ref{sfig:coherences}), separately for the ground-state and the excited-state manifolds, measured in a total of 37 hours. 

\begin{figure}
    \includegraphics[width=\textwidth]{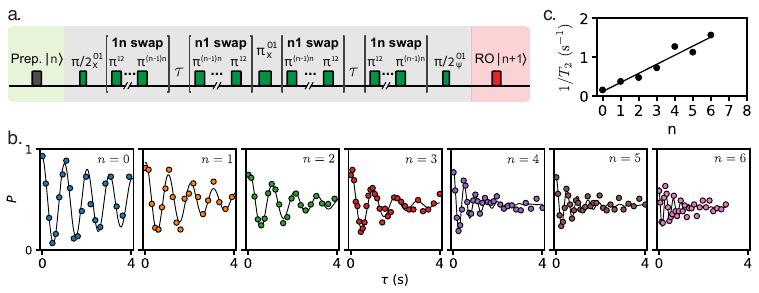}
    \caption{\label{sfig:echo}
    \textbf{Hahn echo for all $\ket{\downarrow 0}$ and $\ket{\downarrow n}$ pairs}.
    \textbf{a.} Pulse sequence. 
    \textbf{b.} Measurement data (dots) is plot alongside an exponentially decaying fit (black solid lines). 
    \textbf{c.} Decoherence rate $1/T_2$ for all $\ket{\downarrow, 0}$ and $\ket{\downarrow, n}$ pairs. Measurements (dots) are fit to a linear trend (black line).
    }
\end{figure}

The large nuclear spin of \Nb offers interesting perspectives for quantum sensing. Indeed, the magnetic moment of states $0$ and $n$ differs by $\sim n \, \mu_\text{\Nb}$, implying that the state $(\ket{0} + \ket{n})/\sqrt{2}$ is $\sim n$ times more sensitive to a small magnetic signal than $(\ket{0} + \ket{1})/\sqrt{2}$. Such "Schrodinger spin-cat" states have been studied in a Sb donor in silicon~\cite{yu_creation_2024}, and used in magnetometry with Rydberg atoms sensing \cite{dietsche_high-sensitivity_2019}. We perform a preliminary characterization of the magnetic sensitivity of these states by measuring the echo coherence of the state $\ket{\downarrow \, 0} + \ket{\downarrow \, n+1}$ (see Fig.\ref{sfig:echo}). The decoherence rate $1/T_2$ is seen to increase linearly with $n$, indicating that decoherence occurs mainly because of magnetic noise. These measurements were taken in a different magnetic field configuration at an angle $\theta=-0.3^\circ$. 

In all measurements, the final $\pi/2$-pulse was applied with a phase linearly increasing with time. In the case of Ramsey and excited state Ramsey, this phase is chosen to nearly compensate the natural oscillation of the signal. In the case of the Hahn echoes, it facilitates fitting the decay time constant.

\section{Generalized correlated echo sequence and differential frequency extraction}
\label{app:correlation_echo}

In this section we describe the generalized correlated echo sequence used to extract differential transition frequencies for the ladder manifold ${|n\rangle}$. In the main text we present the case $n=0$; here we show the full sequence and measurements for all adjacent triplets $n,n+1,n+2$.

We consider a ladder of eigenstates with energies $E_n$ and transition frequencies defined as $\omega_{n,n+1} = (E_{n+1}-E_n)/\hbar$. The pulse sequence begins in $|n\rangle$. A $\pi/2$ pulse on the $n\leftrightarrow n+1$ transition prepares the superposition state $(|n\rangle + |n+1\rangle)/\sqrt{2}$. During the first free evolution interval of duration $\tau$, a relative phase accumulates at frequency $\omega_{n,n+1}$, yielding the state $(|n\rangle + e^{-i\omega_{n,n+1}\tau}|n+1\rangle)/\sqrt{2}$.

The three-pulse block $\pi_{n,n+1},\pi_{n+1,n+2},\pi_{n,n+1}$ coherently maps the population according to $|n\rangle \rightarrow |n+1\rangle$ and $|n+1\rangle \rightarrow |n+2\rangle$ (up to global phases). Importantly, this block reverses the phase accumulated in the first evolution interval, analogous to a spin echo. After this mapping, the state becomes $(|n+1\rangle + e^{+i\omega_{n,n+1}\tau}|n+2\rangle)/\sqrt{2}$.

During the second free evolution interval of duration $\tau$, the relative phase evolves at frequency $\omega_{n+1,n+2}$. The total accumulated phase at the end of the sequence is therefore $\phi(\tau) = (\omega_{n+1,n+2} - \omega_{n,n+1})\tau$. A final $\pi/2$ pulse on the $n+1\leftrightarrow n+2$ transition converts this phase into a population difference. The measured probability oscillates as $P_{|n+1\rangle}(\tau) = \frac{1}{2}[1 + \cos((\omega_{n+1,n+2}-\omega_{n,n+1})\tau)]$.

Thus, the correlated echo directly measures the differential frequency
$\Delta\omega^{(n)} = \omega_{n+1,n+2} - \omega_{n,n+1}$.
The experimentally extracted values for all $n=0,\dots,7$ are summarized in Table~\ref{tab:delta_omega}. 
In a perfectly harmonic ladder, where $\omega_{n+1,n+2} = \omega_{n,n+1}$, 
the correlation signal vanishes. A non-zero correlation frequency therefore directly probes spectral anharmonicity induced by quadrupole and higher-order multipole interactions.

\begin{figure}
    \includegraphics[width=\textwidth]{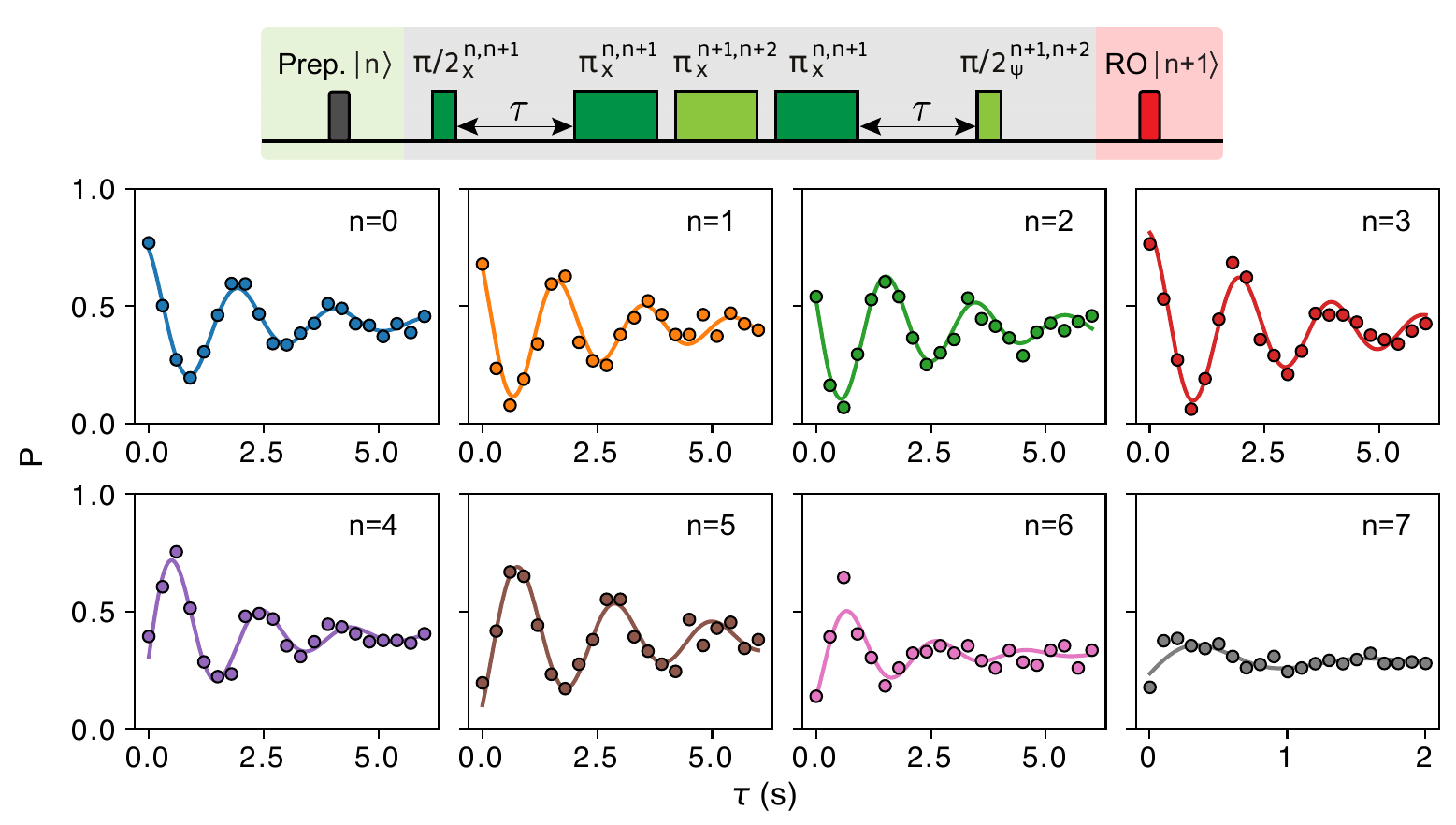}
    \caption{\label{sfig:All_correlation_echo}
    \textbf{Generalized correlated echo sequence and differential frequency measurements.}.
    Top: Pulse sequence used to extract the differential transition frequency $\omega_{n+1,n+2}-\omega_{n,n+1}$. The sequence prepares a superposition on the $n\leftrightarrow n+1$ transition, maps the accumulated phase to the $n+1\leftrightarrow n+2$ manifold using a $\pi_{n,n+1}\pi_{n+1,n+2}\pi_{n,n+1}$ block, and converts the resulting phase into population via a final $\pi/2$ pulse.
    Bottom: Measured correlated echo signals for $n=0,\dots,7$. The oscillation frequency corresponds to $\omega_{n+1,n+2}-\omega_{n,n+1}$.
    }
\end{figure}

\begin{table}[ht]
\centering
\caption{Measured differential transition frequencies 
$\Delta\omega^{(n)} = \omega_{n+1,n+2} - \omega_{n,n+1}$ 
extracted from correlated echo measurements (Fig.18). 
Uncertainties are $1\sigma$ fit errors.}
\label{tab:delta_omega}
\begin{tabular}{cc}
\hline
$n$ & $\Delta\omega^{(n)}$ (Hz) \\
\hline
0 & $-665\,817.024(5)$ \\
1 & $-667\,285.451(7)$ \\
2 & $-669\,699.820(7)$ \\
3 & $-673\,897.780(6)$ \\
4 & $-681\,733.070(10)$ \\
5 & $-697\,995.572(7)$ \\
6 & $-729\,528.188(25)$ \\
7 & $-952\,113.738(31)$ \\
\hline
\end{tabular}
\end{table}

\section{Extraction of frequency uncertainties via bootstrapping}
\label{app:bootstrapping}

Uncertainties of the Ramsey and correlated echo frequencies are estimated using a bootstrap procedure and propagated to the Hamiltonian fits.

\begin{figure}
    \includegraphics[width=\textwidth]{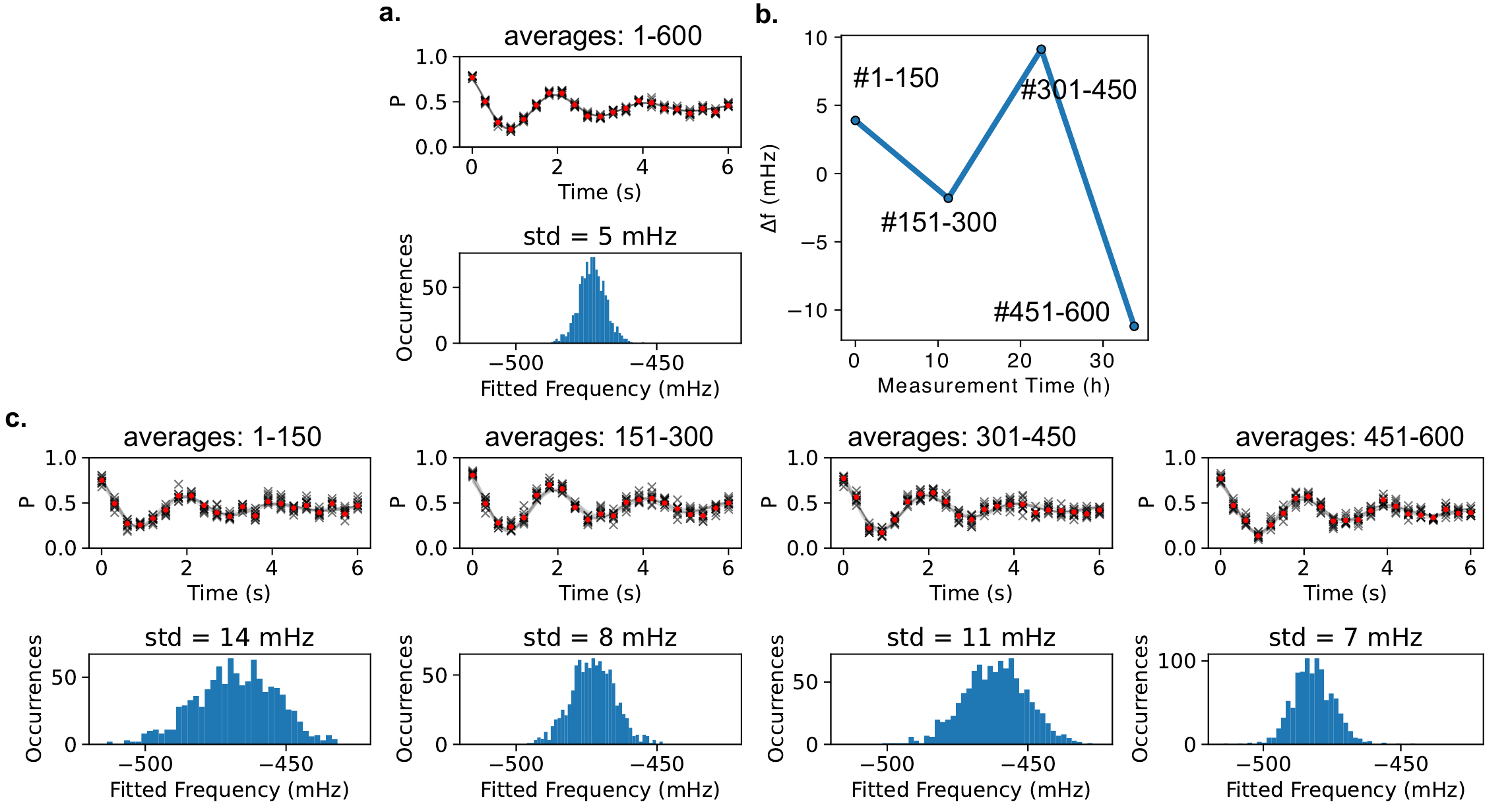}
    \caption{\label{sfig:bootstrapping}
    \textbf{Extraction of frequency uncertainties using bootstrapping.}.
    \textbf{a.} Bootstrap analysis of the first correlated echo transition measured over $\sim$30 hours (600 averages). Photon counts are resampled 1000 times at each echo time (black crosses), and each realization is independently fitted. The distribution of fitted frequencies (bottom) yields a standard deviation of 5 mHz.
    \textbf{b.} Fitted frequencies obtained from four consecutive subsets of 150 averages, showing a standard deviation of $\sim$10 mHz.
    \textbf{c.} Bootstrap analysis applied to each subset, yielding an average uncertainty of $\sim$10 mHz, consistent with $\sqrt{N}$ scaling.
    }
\end{figure}

We consider the first transition of the correlated echo measurement acquired over $\sim$30 hours with 600 averages. For each echo time, photon counts recorded with the single microwave photon detector are resampled 1000 times from the full dataset. Each resampled dataset (black crosses in Fig.\ref{sfig:bootstrapping}a) is independently fitted to extract the transition frequency. The red markers denote the mean over all 600 averages. The distribution of fitted frequencies (Fig.\ref{sfig:bootstrapping}a, bottom) yields a standard deviation of 5 mHz, which we take as the statistical uncertainty.

To validate this estimate, the dataset is divided into four consecutive subsets of 150 averages, corresponding to the acquisition conditions of the remaining correlated echo measurements. The fitted frequencies of these subsets are shown in Fig.\ref{sfig:bootstrapping}b and exhibit a standard deviation of $\sim$10 mHz due to slow drift during the measurement. Applying the same bootstrap procedure to each subset (Fig.\ref{sfig:bootstrapping}c) yields an average uncertainty of $\sim$10 mHz. The increase from 5 mHz to 10 mHz is consistent with the expected $\sqrt{N}$ scaling for four times fewer averages. This agreement confirms the validity of the bootstrap-derived uncertainties used in the Hamiltonian fitting.

\section{Hamiltonian fitting procedure}
\label{app:quadrupole_fitting_procedure}

In this section we detail the procedure for the extraction of the quadrupole and hexadecapole parameters via Hamiltonian fitting.

\subsection{Quadrupole fit for ground and excited state independently}

First we describe the independent fits for $\mathcal{H}^{(\uparrow)}$ and $\mathcal{H}^{(\uparrow)}$, followed by the generalization of the procedure for the full Hamiltonian fit including the SDQ term. Finally, we use the same method to extract the hexadecapole term.

As detailed on the text, the ground and excited state Hamiltonians are given by

\begin{align}
    \mathcal{H} = &\ket{\downarrow}\bra{\downarrow}\otimes \mathcal{H}^{(\downarrow)} + \ket{\uparrow}\bra{\uparrow}\otimes \mathcal{H}^{(\uparrow)} \\
    \label{eq:Hdown}
    \mathcal{H}^{(\downarrow)}/\hbar &= -\omega_S / 2 + \omega_I^{(\downarrow)} I_z^{(\downarrow)} + \mathbf{I^{(\downarrow)}}\cdot \bar{\bar{Q}}^{(\downarrow)}\cdot \mathbf{I^{(\downarrow)}} \\ 
    \label{eq:Hup}
    \mathcal{H}^{(\uparrow)}/\hbar &= +\omega_S / 2 + \omega_I^{(\uparrow)}I_z^{(\uparrow)} + \mathbf{I^{(\uparrow)}}\cdot \bar{\bar{Q}}^{(\uparrow)} \cdot\mathbf{I^{(\uparrow)}}.
\end{align}

\noindent we now focus on the ground state Hamiltonian $\mathcal{H}^{(\downarrow)}$ but the derivation is identical for the excited state Hamiltonian. The variables to fit are $B^{(\downarrow)}_0 = - \omega_I^{(\downarrow)} / \gamma_\text{\Nb}$ and $\bar{\bar{Q}}^{(\downarrow)}$, where $B^{(\downarrow)}_0$ is the effective magnetic field the \Nb spin observes when the \Er is in the ground state, which combines the external field $B_0$ and the effective field generated by the \Er. As demonstrated in App.\ref{app:spin_hamiltonian}, the full quadrupole tensor cannot be measured in a single axis measurement, since a rotation of the quadrupole along the quantization axis leaves the spectrum unchanged. When parameterizing the quadrupole in terms of the rank-2 spherical tensors (see App.\ref{app:spin_hamiltonian}), the set of variables to fit is defined as $\mathcal{V^{(\downarrow)}} \equiv\{B^{(\downarrow)}_0, S^{(\downarrow)}_0, S^{(\downarrow)}_1, S^{(\downarrow)}_2, \Delta^{(\downarrow)}\}$ and $\zeta$ is arbitrarily set to 0. Another degeneracy inherent to the one-axis nature of the measurement is the sign of $S_1^{(\downarrow)}$ and $S_2^{(\downarrow)}$ as well as $\Delta^{(\downarrow)}$, that is periodic between $-\pi/4$ and $\pi/4$. For the purposes of the fit, we set the following priors

\begin{equation}
    S_1^{(\downarrow)} > 0 \quad | \quad S_2^{(\downarrow)}>0 \quad | \quad -\pi/4 <\Delta^{(\downarrow)} < \pi/4
\end{equation}

The ground state NMR transitions $\omega_{n,\ n+1}^{\text{fit}}(\mathcal{V^{(\downarrow)}})$ are calculated by numerically diagonalizing $H^{(\downarrow)}(\mathcal{V}^{(\downarrow)})$ and taking the energy differences between the relevant energy levels. We define the following likelihood

\begin{equation}
    \mathcal{L}^{(\downarrow)}(\mathcal{V}^{(\downarrow)}) = \sum_{n=0}^9 \left(\frac{\omega^{(\downarrow)}_{n, \ n+1} - \omega_{n, \ n+1}^{\text{fit}}(\mathcal{V^{(\downarrow)}})}{\sigma^{(\downarrow)}_{n, \ n+1}}\right)^2,
\end{equation}

\noindent where $\sigma^{(\downarrow)}_n=1$~Hz is the measurement uncertainty and $\omega^{(\downarrow)}_n$ correspond to the measured NMR transition frequencies in the ground state,
 
\begin{align}
\omega^{(\downarrow)}_{n,n+1}/2\pi =
\begin{aligned}[t]
(&7560562.0(2),\;
6894745.1(2),\;
6227459.8(2),\\
&5557759.9(2),\;
4883861.9(2),\;
4202128.8(1),\\
&3504133.1(2),\;
2774604.8(2),\;
1822491.2(2))\ \mathrm{Hz}
\end{aligned}
\end{align}

given here in increasing order of $n$. Respectively, the NMR transitions for the excited state are the following

\begin{align}
\omega^{(\uparrow)}_{n,n+1}/2\pi =
\begin{aligned}[t]
(&-136547(45),\;
-135922(30),\;
-135196(24),\\
&-134203(28),\;
-132832(22),\;
-130986(18),\\
&-128470(16),\;
-122551(20),\;
-128757(18))\ \mathrm{Hz}
\end{aligned}
\end{align}

The Hamiltonian fit was performed via Monte-Carlo Markov-Chain simulation \cite{foreman-mackey_emcee_2013}. In brief, the algorithm initializes an arbitrary number of walkers, or particles, in the parameter space. After an evaluation of the logarithm of the likelihood function, the positions of the walkers are updated through the "stretch move" \cite{goodman_ensemble_2010}. We run the algorithm with 64 walkers and 60000 iterations, from which we obtain the following median values for the ground and excited state parameters

\begin{align}
\begin{split}
B^{(\downarrow)}_0 &= 460.54333(8)\,\text{mT}, \\
S^{(\downarrow)}_0 &= -237.3530(1)\,\text{kHz}, \\
S^{(\downarrow)}_1 &= 2.667(8)\,\text{kHz}, \\
S^{(\downarrow)}_2 &= 149.443(1)\,\text{kHz}, \\
\Delta^{(\downarrow)} &= -0.002(60),
\end{split}
\qquad
\begin{split}
B^{(\uparrow)}_0 &= 447.732(7)\,\text{mT}, \\
S^{(\uparrow)}_0 &= -237.224(11)\,\text{kHz}, \\
S^{(\uparrow)}_1 &= 6.1(3)\,\text{kHz}, \\
S^{(\uparrow)}_2 &= 149.48(8)\,\text{kHz}, \\
\Delta^{(\uparrow)} &= -0.01(24).
\end{split}
\end{align}

We calculate the reduced chi-squared for the ground and excited state and obtain $\chi^{2,(\downarrow)}_\nu=0.75$ and $\chi^{2,(\uparrow)}_\nu=1.14$. The proximity of these values to 1 indicates that the model correctly describes the data while avoiding over-fitting. We remark that the uncertainty of $S_1^{(\downarrow)}$ and $\Delta^{(\downarrow)}$ is much larger than the other parameters. This is readily explained by the effect of these quantities on the energy levels of the Hamiltonian. On the one hand, $S_1$, $\Delta$ are closely related to the values of the quadrupole perpendicular to $B_0$, and are only observable as second and third order effects respectively. On the other hand, $B_0$, $S_0$ and $S_2$ are responsible for zero and first order effects and their estimation is more precise. 

\begin{figure}[t]
    \includegraphics[width=\textwidth]{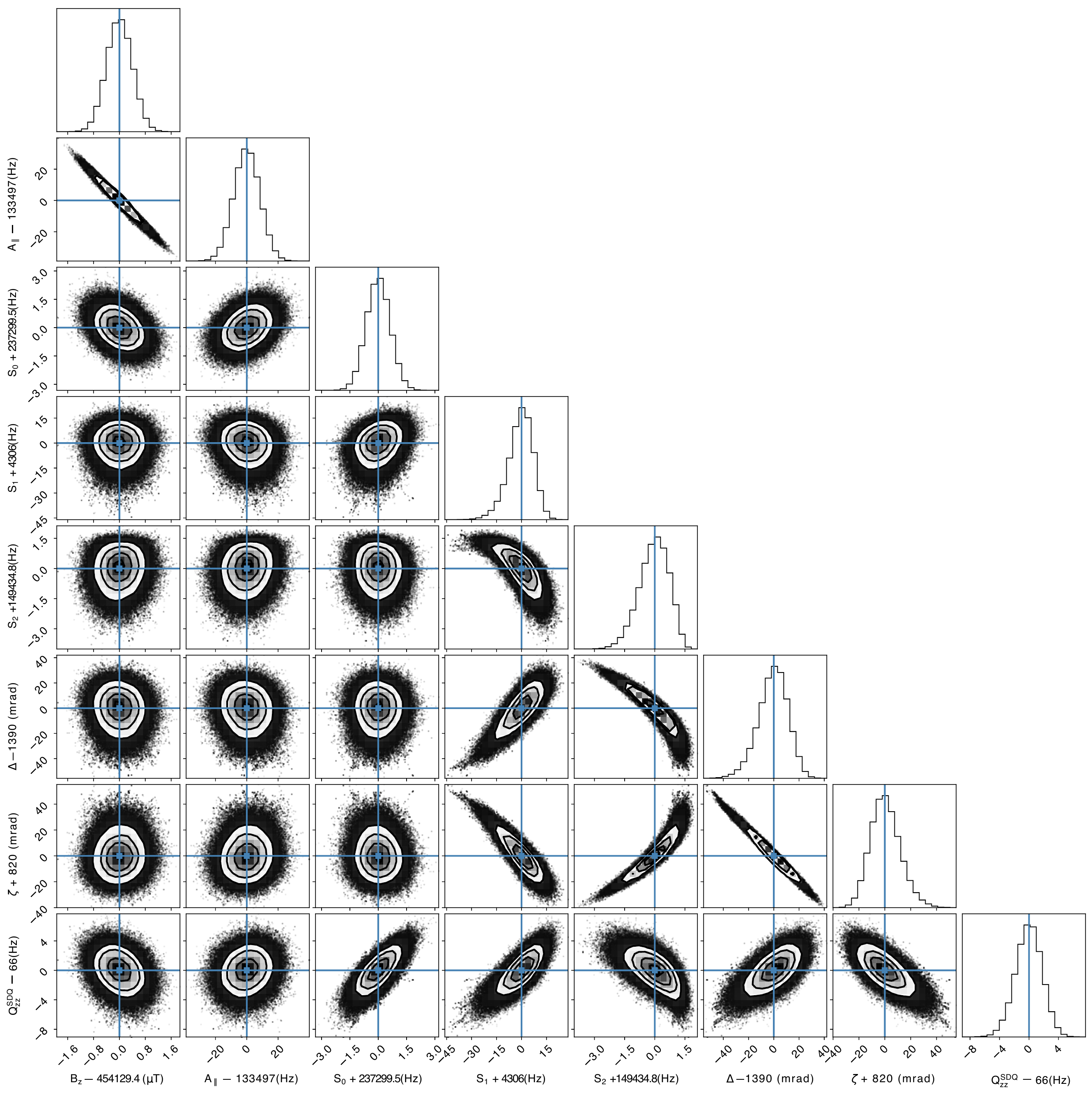}
    \caption{\label{sfig:corner_plot}
    \textbf{Corner plots of the posterior distributions of the full Hamiltonian fit}. Diagonal plots convey the one-dimensional distribution of the samples and off-diagonal plots the two-dimensional projection between two parameters. Vertical lines show the median value of the parameters, which we take as the result of the fit.
    }
\end{figure}

The posterior distributions of the fits, obtained with the last 5000 iterations, are plot in Fig.\ref{sfig:corner_plot} in the form of a corner plot \cite{foreman-mackey_cornerpy_2016}. The high uncertainty of $S_1$ and $\Delta$ yields strong non-linear correlations, which are visualized on the two-dimensional plots as strong non-gaussian behavior.



\subsection{Quadrupole fit for the full Hamiltonian}

We now extend this procedure for the full Hamiltonian fit. As detailed in the text, the presence of a perpendicular hyperfine coupling results in an effective two-axis measurement. This allows to completely fit the quadrupole tensor in the reference frame defined by the hyperfine coupling. As detailed in App.\ref{app:spin_hamiltonian}, the full spin-1/2 Hamiltonian is given by   

\begin{align}
\begin{split}
        \mathcal{H}/\hbar &= \omega_S\cdot S_z + \omega_I\cdot I_z +\mathbf{S} \cdot \bar{\bar{A}} \cdot \mathbf{I} + \mathbf{I} \cdot \bar{\bar{Q}} \cdot \mathbf{I} + Q_\mathrm{sdq}\, S_z \tfrac12[3 I_z^2 - I(I+1) ]  + \frac{g_0^2\left(\omega_0 - \omega_n\right)}{\left(\omega_0 - \omega_n\right)^2 + \kappa^2/4} \ket{\uparrow,n}\bra{\uparrow,n}
\end{split}
\end{align}


We consider $A_{zx}=A_\perp=55(8)$~kHz to be a fixed parameter, as it is uniquely determined by the measurements presented in App.\ref{app:stimulated_raman_driving}. In addition, we fix the non-secular terms to those calculated using eq. \ref{eq:full-dipole-dipole-hamiltonian} to avoid overfitting. Following the parameterization presented above, the remaining set of eight variables to fit is defined as $\mathcal{V}\equiv\{B_0, A_\parallel, S_0,S_1,S_2,\Delta,\zeta,Q_\mathrm{sdq}\}$. Naturally, we define the total likelihood as $\mathcal{L}(\mathcal{V}) = \mathcal{L}^{(\downarrow)}(\mathcal{V}) + \mathcal{L}^{(\uparrow)}(\mathcal{V})$ and fit the data with the same Monte-Carlo Markov-Chain algorithm with the "stretch move". As before, convergence of this fit was confirmed via visual inspection of the walker's trace. The median values of the with 64 particles and 60000 iterations are

\begin{align}
\label{seq:full_fit_results}
\begin{split}
    B_0 =  454.129(1)\text{~mT} \; &| \; A_\parallel =  133.497(8)\text{~kHz} \\ 
    S_0 = -237.299(1)\text{~kHz} \; &| \; S_1 = -4.36(1) \text{~kHz} \\
    S_2 = -149.435(1)\text{~kHz}\; &| \;  Q_\mathrm{sdq} = 66(6) \text{~Hz}\\
    \Delta = 1.39(1)\; &| \;\zeta = -0.82(1) \\
\end{split}
\end{align}

The reduced chi-squared of the model is $\chi^2_\nu=1.2$. The uncertainty of the fit has two components: the standard deviation of the posterior distributions shown in Fig.\ref{sfig:corner_plot} and the standard deviation of the fit results when performing the MCMC fit with 400 randomly sampled values of $A_\perp$, assuming a Gaussian distribution for $A_\perp$ with a standard deviation of 9~kHz. 

From these values we reconstruct the nuclear quadrupole tensor $\bar{\bar{Q}}$ and diagonalize it to obtain the principal axis of the quadrupole $X$, $Y$ and $Z$ and the corresponding strengths $Q_{X} = 268.138(9)$~kHz, $Q_{Y} = -30.799(7)$~kHz and $Q_{Z} = -237.338(3)$~kHz.  We find that the $Z$-axis makes an angle of 0.06(3) with respect to the $c$-axis and the $X$-axis makes and angle of -10(6) with respect to the $a$-axis of the crystal.


We calculate the quadrupole moment $C_q=19.3059(7)$~MHz and the anisotropy $\eta=0.77027(7)$. These values, as well as the principal axis orientation, quantitatively agree with the DFT calculations performed for a system with the same crystal but with an yttrium atom (see App.\ref{app:DFT_calculation}). Figure \ref{sfig:corner_plot} plots the posterior distribution of the last 5000 iterations for $\mathcal{V}$. We note the reduction of the inter-parameter correlations, which we attribute to the two-axis nature of the measurement, which reduces uncertainty in perpendicular components of the quadrupole. This is evidenced in the reduction by an order of magnitude in the uncertainty of the $S_1$ and $\Delta$ parameters compared to the results of the ground and excited state fits.




\begin{figure}[t]
    \includegraphics[width=0.8\textwidth]{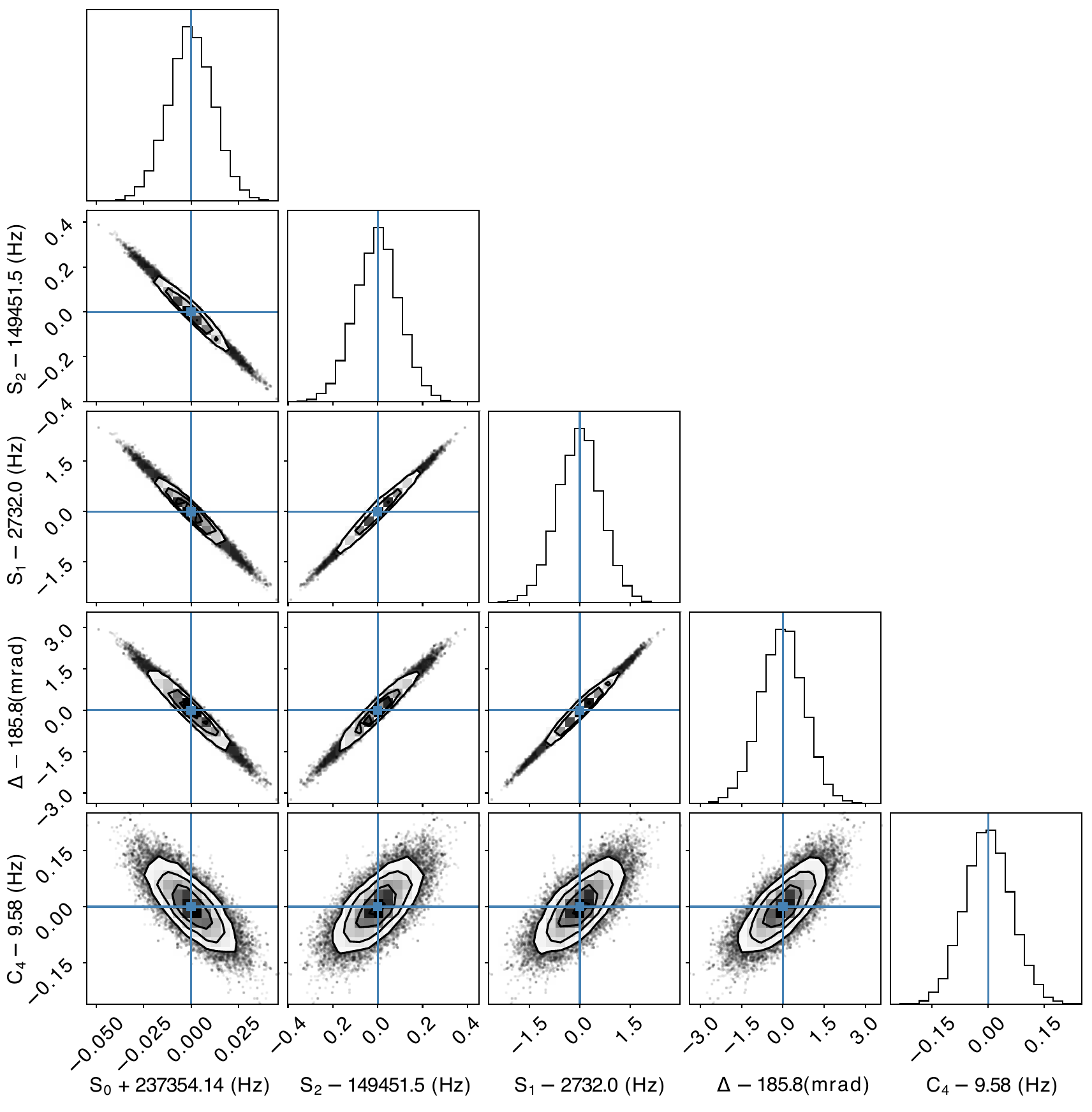}
    \caption{\label{sfig:hex_corner_plot}
    \textbf{Corner plots of the posterior distributions of the Hamiltonian fit for extracting the hexadecapole term}. Diagonal plots convey the one-dimensional distribution of the samples and off-diagonal plots the two-dimensional projection between two parameters. Vertical lines show the median value of the parameters, which we take as the result of the fit.
    }
\end{figure}    

\subsection{Hexadecapole fit}

The method used to extract the hexadecapole coupling strength $C_4$ is similar to the quadrupole fitting procedure. In particular, we fit the differential frequencies in Table.\ref{tab:delta_omega} to the effective nuclear-spin Hamiltonian

\begin{align}
    \label{eq:hex}
    \mathcal{H}^{(\downarrow)}/\hbar &= -\omega_S / 2 + \omega_I^{(\downarrow)} I_z^{(\downarrow)} + \mathbf{I^{(\downarrow)}}\cdot \bar{\bar{Q}}^{(\downarrow)}\cdot \mathbf{I^{(\downarrow)}} + \tfrac{C_4}{I(2I-1)(I-1)(2 I -3)}  I_z^4.
\end{align}

The posterior distribution is shown in Fig.\ref{sfig:hex_corner_plot}, with the following fitted parameters

\begin{equation}
\label{seq:full_fit_results}
\begin{aligned}
S_0 &= -237.35414(1)\,\text{kHz} \\
S_2 &= 149.4515(1)\,\text{kHz} \\
S_1 &= 2.7320(7)\,\text{kHz} \\
\Delta &= 0.1858(8) \\
C_4 &= 9.6(1)\,\text{Hz}
\end{aligned}
\end{equation}

\section{Pseudo-quadrupole interaction}
\label{app:pseudo_quadrupole_interaction}

\begin{figure}[t]
    \includegraphics[width=\textwidth/2]{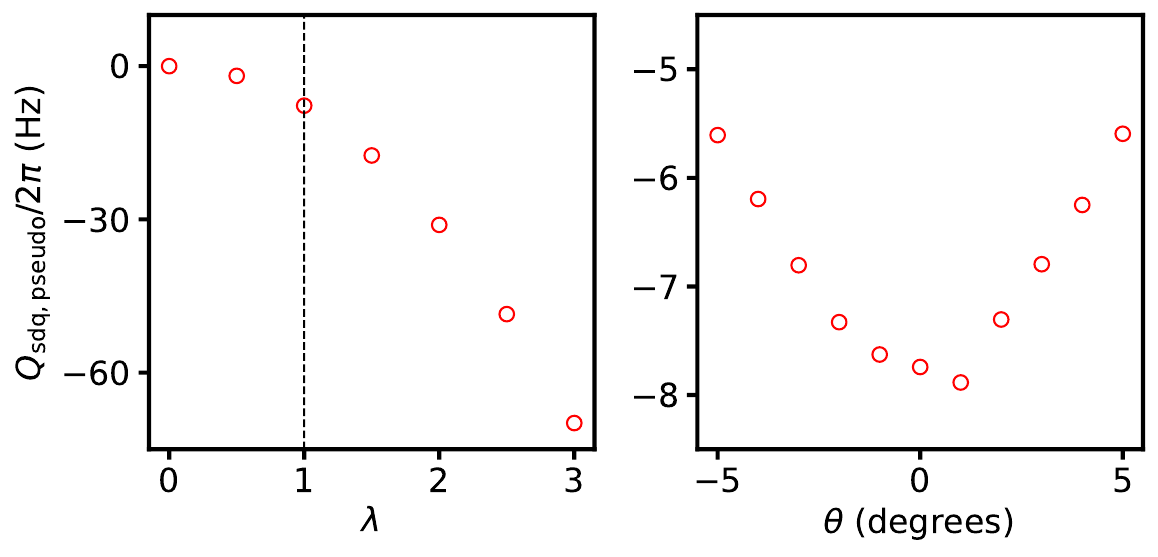}
    \caption{\label{sfig:pseudoquadrupole}
    \textbf{Pseudo-quadrupole simulation.}
    \textit{(left)} Effective SDQ interaction from the pseudo-quadrupole, $Q_\mathrm{sdq, pq}$, is plot as a function of the hyperfine scaling parameter $\lambda$. The vertical line marks $\lambda=0$. The simulation was performed for $\theta=-0.5^\circ$ and $\beta_0=-0.57^\circ $.
    \textit{(right)} $Q_\mathrm{sdq, pq}$ as a function of the in-plane angle $\theta$ with $\lambda=1$ and $\beta_0=-0.57^\circ$.
    }
\end{figure}

In this section we attempt to explain the observed SDQ effect through the action of higher energy Kramers doublets $Z_2, \dots,Z_8$ on the nuclear transition energies of the \Nb spin. As explained in App.\ref{app:spin_hamiltonian}, the spin-1/2 model (Eq.~\ref{eq:spin_12_hamiltonian}) is a restriction of the more complete Hamiltonian ((Eq.~\ref{eq:spin_complete_ham}) to the lowest two \Er energy levels. In particular, the hyperfine Hamiltonian coupling the effective spin $S$ and the nuclear spin is a two-level restriction of the hyperfine Hamiltonian $\mathcal{H}_{hf,J}$ (Eq.C17). The latter couples the nuclear spin not only to the $Z_1$ doublet, but also to the other higher-energy $Z_{2-8}$ states (see Fig.\ref{sfig:energy-levels}). These couplings renormalize the nuclear energy levels in a manner that may mimic an effective quadrupolar interaction. This pseudo-quadrupolar interaction is of purely magnetic origin, contrary to the real quadrupolar interaction\cite{foley_second-order_1947, abragam_electron_2012}, and can be of simiar order of magnitude as the electric quadrupole when the nuclear spin is located on the nucleus of the paramagnetic ion~\cite{schweiger_principles_2001}. 

The effective Hamiltonian of the pseudo-quadrupolar interaction depends on the point-symmetry of the site where the ion is located. It has been predicted that this interaction can be described by a term of the form $S_z I_z^2$ for D$_{3\text{h}}$ symmetry \cite{ghatikar_zeeman-field-dependent_1966}, which would imply that the pseudo-quadrupolar part may depend on the spin state, as observed in our experiment. However, no prediction was made for the $S_4$ symmetry. We therefore rely on numerical simulations to estimate the relevance of this effect.

To quantitatively characterize the effect of the pseudo-Quadrupole for this specific configuration, we proceed by performing the following numerical analysis. First, we generate a series of the NMR transitions for the \Nb -- \Er system by diagonalizing the $J=15/2$ Hamiltonian

\begin{align}
        \mathcal{H} =\ \mu_B \ g_J\mathbf{B_0}\cdot \mathbf{J} + \mathcal{H}_{\mathrm{cf}} 
        + \omega_I I_z + \mathbf{I}\cdot \bar{\bar{Q}}\cdot \mathbf{I}  
    + \ \mathbf{J}\cdot \lambda \bar{\bar{A}}_J\cdot\mathbf{I} +\mathcal{H}_{\mathrm{L}}.
\end{align}

\noindent The hyperfine coupling tensor $\bar{\bar{A}}_J$ is calculated from a dipole-dipole interaction between the \Er spin and the \Nb at the expected position (see App.\ref{app:spin_hamiltonian}). We use the $S_4$-symmetric crystal-field Hamiltonian $\mathcal{H}_{\mathrm{cf}} $ determined in~\cite{enrique_optical_1971}, whose validity is questionable here due to the $S_4$ symmetry breaking (see App.C). Given that the pseudo-quadrupole effect is a second order interaction of the hyperfine coupling, we use a dimensionless scaling parameter $\lambda$ for the hyperfine interaction strength. Otherwise, the Hamiltonian parameters are set to the measured values. 

The NMR transition frequencies computed in the lowest two \Er manifolds are then fitted with the effective spin-1/2 model, using the same procedure (outlined in App.\ref{app:quadrupole_fitting_procedure}) as the one used for the experimental data, yielding the effective quadrupolar interaction in each \Er state. The difference between the principal values along the $c$ axis between the $\uparrow$ and $\downarrow$ values, $Q_\mathrm{sdq, pq}$, is shown in Fig.\ref{sfig:pseudoquadrupole} as a function of the scaling parameter $\lambda$.

We first note that the computed $Q_\mathrm{sdq, pq}$ is $0$ when $\lambda=0$, as expected. For non-zero values of $\lambda$, a non-zero spin-dependent pseudo-quadrupole is computed. Its value scales quadratically with the hyperfine scaling factor, as expected for an effect that can be described in second-order perturbation theory. The magnitude of the interaction is of similar order as the experimentally measured value, although the $\lambda=1$ value $Q_\mathrm{sdq, pq}/2\pi =-8$~Hz is a factor $10$ smaller than the measurement and has an opposite sign. We also compute $Q_\mathrm{sdq, pq}$ at $\lambda=1$ as a function of $\theta$. In a window of $\pm5$ degrees, $Q_\mathrm{sdq, pq}$ remains consistently smaller than 10~Hz in absolute value.

We conclude that pseudo-quadrupole interactions arising from second-order paramagnetic coupling between the electron $J$-multiplet and the nuclear spin needs to be considered in a complete analysis. However, it has opposite sign and is one order of magnitude smaller than the measurements, implying that the effect is likely not the only one at play.

\section{Pseudo-hexadecapole interaction}
\label{app:pseudo_hexadecapole_interaction}

In the measurements, only nuclear transitions within the \Er ground state manifold are probed. However, higher electron states may renormalize the effective nuclear Hamiltonian and mimic higher-order multipole terms. To assess whether the observed hexadecapole term could arise as a \textit{pseudo-hexadecapole} from electronics excited states of \Er, we perform numerical calculations.

Here we consider the \Er excited state. Similar to the pseudo-quadrupole calculations in App.\ref{app:pseudo_quadrupole_interaction}, we introduce a scaling parameter $\lambda$ that multiplies the hyperfine interaction $\lambda \mathbf{J}\cdot \mathbf{A}\cdot\mathbf{I}$, allowing us to continuously tune the electron-nuclear mixing strength. The physically relavant case corresponds to $\lambda = 1$. For each value of $\lambda$, we compute synthetic "measurement" data in the form of differential transition frequencies (see App.\ref{app:correlation_echo}), including Zeeman, quadrupole, and hyperfine, and subsequently fit these data with an effective nuclear Hamiltonian that includes a hexadecapole term. Note that we have to consider the magnetic field as a fitting parameter, in order to accommodate for the $\lambda$-dependent effective Zeeman shift caused by the hyperfine interaction.  

\begin{figure}[t]
    \includegraphics[width=\textwidth]{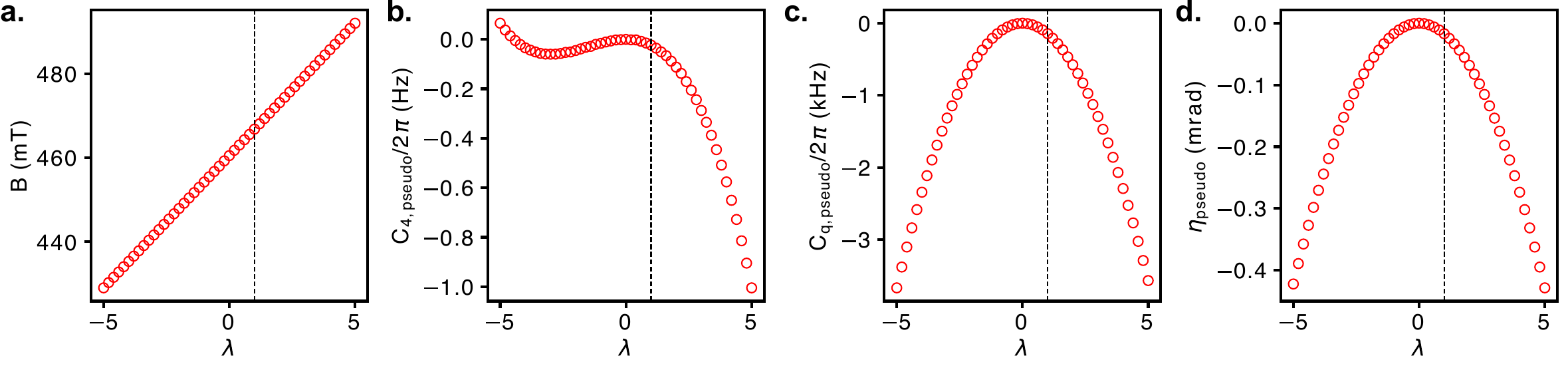}
    \caption{\label{sfig:pseudohexadecapole}
    \textbf{Pseudo-hexadecapole contributions from \Er excited states.}
    \textit{(a)} The fitted magnetic field shifts approximately linearly with $\lambda$, indicating that the fit absorbs most of the hyperfine-induced shifts through a renormalization of the field.
    \textbf{(b)} Extracted pseudo-hexadecapole coupling strength $C_{4,pseudo}/2\pi$ as a function of the hyperfine scaling parameter $\lambda$. At $\lambda = 1$ $C_{4,pseudo}/2\pi$ reaches -0.03 Hz.
    \textbf{(c)} Extracted pseudo-quadrupole moment $C_{q,pseudo}/2\pi$ as a function of the hyperfine scaling parameter $\lambda$.
    \textbf{(d)} Extracted pseudo-quadrupole anisotropy $\eta_{pseudo}/2\pi$ as a function of the hyperfine scaling parameter $\lambda$.
    }
\end{figure}

Results are shown in Fig.\ref{sfig:pseudohexadecapole}
. The fitted magnetic field shifts approximately linearly with $\lambda$ (Fig.\ref{sfig:pseudohexadecapole}a), as expected. The difference between the fitted quadrupole parameters $C_q$ and $\eta$ and the values of the quadrupolar Hamiltonian show a quadratic dependence on $\lambda$ (Fig.\ref{sfig:pseudohexadecapole}c,d), corresponding to the pseudo-quadrupole effect, which is therefore small in our conditions. The fitted hexadecapole term reaches -0.03 Hz at $\lambda = 1$ (Fig.\ref{sfig:pseudohexadecapole}b). This pseudo-hexadecapole contribution is two orders of magnitude smaller than the value obtained from the measurements, and has moreover the opposite sign.  At $\lambda = 0$, both pseudo-quadrupole and pseudo-hexadecapole vanish, as expected since the nuclear spin is completely decoupled from the electronic degrees of freedom.

\section{Electric-dipole calculation}
\label{app:electric-dipole_calculation}

In this section we discuss the strength of the electric dipole generated by the \Er spin and quantitatively estimate the influence of the dipole field on the \Nb quadrupole. 

Historically, the first observations of the magnetically-induced electric dipole were performed by measuring shifts of the EPR transitions under the application of static electric fields. \cite{ham_linear_1961, ludwig_splitting_1961, mims_electric_1964}. The effect of a static electric field on the effective spin-1/2 ground state of Kramers ion in a non-centrosymmetric site can be described with the following Hamiltonian \cite{kiel_theory_1966}


\begin{equation}
    \mathcal{H}_{\text{E}} = \sum_{i,j,k} T_{k, i, j} S_{i} B_j E_k
\end{equation}

where $i,j,k \in\{x,y,z\}$ and $T_{k, i, j}$ is a third-rank tensor which is symmetric in the {i, j} indices. The experimentally determined values for the tensor in \Er:\Ca are given by \cite{mims_electric_1965} 

\begin{equation}
    T_{x,z,x} = 1.4 \quad | \quad T_{x,z,y} = 1.0  \quad | \quad T_{z,x,x} = 5.4 \quad | \quad T_{z,y,x} = 2.9  \quad |\quad \left[10^{-32}   \frac{\text{J/T}}{\text{V/m}} \right]
\end{equation}

\noindent and the rest of the elements of the tensor are obtained by enforcing $S_4$ symmetry and $\{i, j\}$ index symmetry

\begin{equation}
    T_{z,y,y} = - T_{z,x,x} \quad | \quad T_{y,z,y} = - T_{x,z,x} \quad | \quad T_{y,z,x} = - T_{x,z,y} \quad | \quad T_{k,i,j} = T_{k,j,i},
\end{equation}

At a fixed magnetic field, and in the high-field regime ($\omega_S\gg T_{k,i,j}$), the interaction can be rewritten as an effective electric dipole $\mathbf{d}$ under an electric field:

\begin{align}
\begin{split}
    \mathcal{H}_{\text{E}} &= \sum_k-d_k E_k = -\mathbf{d} \cdot \mathbf{E}, \\
    d_k &= - \sum_{i,j} T_{k, i, j} \langle S_{i}\rangle B_j.
\end{split}
\label{seq:electric_dipole}
\end{align}

where $\langle S_i \rangle$ corresponds to the expected value of the electron spin in each of the crystalline principal axis. 

\begin{align}
\begin{split}
    \langle\mathbf{S}\rangle &= \frac12\frac{\bar{\bar{\gamma}} \cdot\mathbf{B}  }{||\bar{\bar{\gamma}} \cdot\mathbf{B}||}
\end{split}
\end{align}

We note that, in the high-field regime, the spin orientation in the ground state and excited state are opposite and the associated electric dipole moments of each state also point in opposite directions. Consequently, the electric field and electric field gradient generated by the dipole will be different depending on the electron spin state. Moreover, electric fields are known to induce linear Stark shifts of the nuclear quadrupole in non-centrosymmetric sites \cite{noauthor_national_1961, armstrong_linear_1961}, arising from ion displacements that modify the local field gradient. For example, the measured sensitivity of the $^{151}$Eu quadrupole in Y$_2$SiO$_5$ ranges between 0.1 and 1~Hz/(V/cm) \cite{macfarlane_optical_2014}.

It is important to note that for \Er:\Ca the dipole moment is strictly zero if the magnetic field is parallel to the $c$-axis. However, during this experiment the field was misaligned with the $(a,c)$-plane of the crystal by an angle $\beta_0=-0.57^\circ$ as well as an angle of $\theta_0=-0.6^\circ$ with respect to the projection of the $c$ axis in the resonator plane (see App.\ref{app:sample_details} and App.\ref{app:electron-spin-characterization}). We estimate the strength of the dipole by evaluating eq. \ref{seq:electric_dipole} and obtain $\mathbf{d} = \pm$(-0.26, 0.25, 0.02)~mD. The dipole is confined in the $(a,b)$-plane of the crystal and generates both an electric field and electric field gradient at the position of the \Nb, with values

\begin{align}
\begin{split}
    &\mathbf{E} = (34,~-33.6,~7) \text{~V/cm}\\
    &\partial E_z / \partial z = V^{\text{(d)}}_{zz} = 0.02~\text{\textmu V/\r{A}}^2
\end{split}
\end{align}

The electric field gradient induces a spin-dependent quadrupole of $Q_\text{sdq, dipole} \approx 0.2$~mHz, far smaller than the observed interaction (76~Hz). The electric field at the Nb site is $\sim$48~V/cm, giving a differential field of $\sim$96~V/cm between the two spin states. If we assume a similar sensitivity for \Nb:\Ca as for the previously studied nuclear spins \cite{macfarlane_optical_2014} we obtain an interaction strength of $\sim$10 -- 100~Hz, consistent with the measured value.

\section{DFT calculation}
\label{app:DFT_calculation}

\begin{figure}[t]
    \includegraphics[width=\textwidth/2]{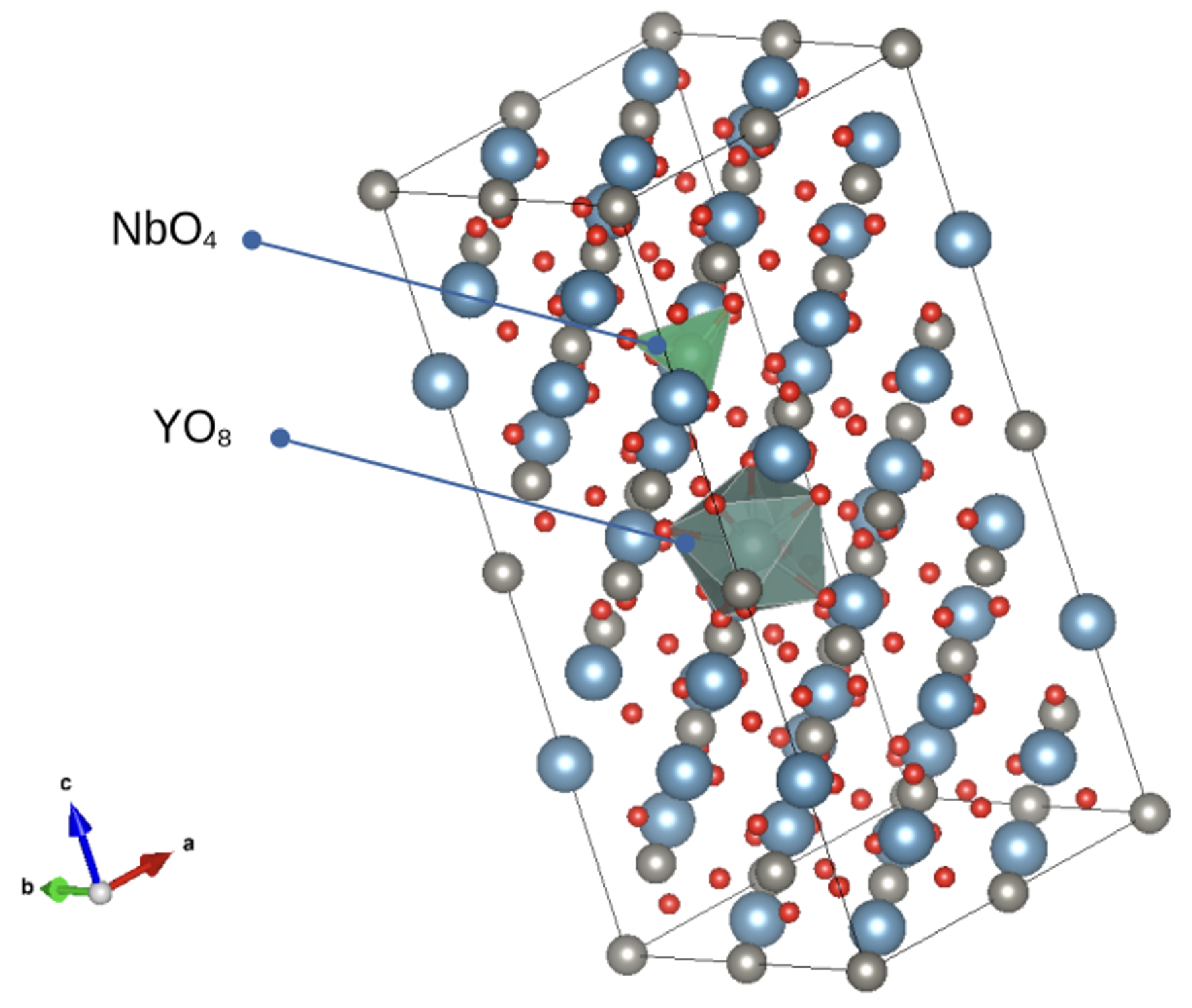}
    \caption{\label{sfig:DFT_calculation}
    \textbf{View of the $(2\times2\times2)$ structural model used to compute the NMR properties (EFG tensor) of \Nb.}
    }
\end{figure}

The DFT-NMR calculations were performed with two codes devised for periodic solids, namely CP2K~\cite{Hutter2014} and VASP~\cite{Kresse1996}, following the methodology developed to study NMR properties of large disordered systems such as glasses (see for example~\cite{Bertani2023}). A supercell $(2\times2\times2)$ of the \Ca was constructed in order to minimize the interactions between \Er and Nb$^{3+}$ and their images under the periodic boundary conditions. Then one W atom was substituted for Nb$^{3+}$, and the adjacent Ca$^{2+}$ atom along the c-axis was substituted for Y$^{3+}$ to model the impact of \Er without the complexity of dealing with the unpaired electron. This lead to a structural model of 192 atoms shown in Fig.~\ref{sfig:DFT_calculation}. The configuration was optimized using the CP2K codes (atomic coordinates and edge lengths of the orthorhombic supercell) with the GGA-DFT PBE (Perdew-Burke-Ernzerhof) functional~\cite{Perdew1996} with dispersion corrections as implemented by the DFTD3 method of Grimme et at.~\cite{Grimme2010} The EFG tensors were then calculated using the PAW method as implemented in VASP~\cite{Vasconcelos2013}. La$^{3+}$ and Lu$^{3+}$ were also investigated but with results in less agreement with the experimental value.

\end{appendix}

\newpage
\bibliographystyle{naturemag}
\bibliography{93NbSpectro}

\end{document}